\documentclass[12pt]{article}
\usepackage{amsfonts}
\newcommand{\be}{\begin{equation}}
\newcommand{\ee}{\end{equation}}
\newcommand{\bea}{\begin{eqnarray}}
\newcommand{\eea}{\end{eqnarray}}
\newcommand{\nn}{\nonumber \\}
\newcommand{\p}[1]{(\ref{#1})}
\newcommand{\lb}{\label}
\def\sfrac#1#2{{\textstyle\frac{#1}{#2}}}

\begin{document}

\begin{titlepage}

\vfill

\begin{flushright}
ITP--UH--23/14, JINR E2--2014--99
\end{flushright}

\vspace{3.0cm}

\begin{center}
{ \bf \LARGE
 Auxiliary tensor fields for Sp(2,\,$\mathbb{R}$) self-duality}\\
\end{center}
\vspace{1.0cm}
\begin{center}
{\Large Evgeny A. Ivanov${\,}^{a)}$, Olaf Lechtenfeld${\,}^{b)}$, Boris M. Zupnik${\,}^{a)}$}\\
\end{center}
\vspace{1.0cm}

\centerline{${\,}^{a)}$\it Joint Institute for Nuclear Research,
Dubna, Moscow Region, 141980 Russia}
\vspace{0.5cm}

\centerline{${\,}^{b)}$ \it Institut f\"ur Theoretische Physik
and Riemann Center for Geometry and Physics,}
\centerline
{\it Leibniz Universit\"at Hannover,
Appelstra{\ss}e 2, 30167 Hannover, Germany}
\vspace{0.5cm}

\centerline{E-mail: eivanov@theor.jinr.ru,~lechtenf@itp.uni-hannover.de,~zupnik@theor.jinr.ru}

\vspace{3.0cm}

\par
\begin{center}
{\bf ABSTRACT}

\end{center}
The coset Sp(2,$\,\mathbb{R}$)$/$U(1) is parametrized by two real
scalar fields. We generalize the formalism of auxiliary tensor
(bispinor) fields in U(1) self-dual nonlinear models of abelian
gauge fields to the case of Sp(2,$\,\mathbb{R}$) self-duality. In
this new formulation, Sp(2,$\,\mathbb{R}$) duality of the nonlinear
scalar-gauge equations of motion is equivalent to an
Sp(2,$\,\mathbb{R}$) invariance of the auxiliary interaction. We
derive this result in two different ways, aiming at its further
application to supersymmetric theories. We also consider an
extension to interactions with higher derivatives.

\vspace{4.0cm}

\end{titlepage}
\setcounter{equation}0
\section{Introduction}

Noncompact Sp(2,$\,\mathbb{R}$) duality arises in nonlinear electrodynamics interacting
with dilaton and axion scalar fields which support a nonlinear realization of  Sp(2,$\,\mathbb{R}$) in the coset
Sp(2,$\,\mathbb{R}$)$/$U(1) \cite{GZ}-\cite{AFZ}. The  Sp(2,$\,\mathbb{R}$) self-dual Lagrangian contains a specific interaction
of the electromagnetic
field ${F}_{mn}$ and the coset scalar fields $S=(S_1, S_2)$ \cite{GR2},
\be
L^{\rm sd}(S, {F}) = L(S)+\sfrac14S_1{F}^{mn}\tilde{{F}}_{mn}+\hat{L}(\sqrt{S_2}{F}_{mn})\,, \lb{Sp2Rlagr}
\ee
where $\hat{L}(F) = -\frac14 F^{mn}F_{mn} + \hat{L}^{\rm int}(F)$ is any Lagrangian of U(1) duality-invariant systems of  nonlinear electrodynamics \cite{GZ,GR}, and $L(S)$
is a sigma-model-type Lagrangian for the coset scalar fields.
Like in the U(1) duality case, the   Sp(2,$\,\mathbb{R}$) is respected by the full system of equations of motion
following from \p{Sp2Rlagr} together with the Bianchi identity for
${F}^{mn}\sim ({F}_{\alpha\beta}, \bar{F}_{\dot\alpha\dot\beta})$, while \p{Sp2Rlagr} on its own does not possess either
 Sp(2,$\,\mathbb{R}$) or U(1) invariance. Coset-field-extended self-dual systems of abelian gauge fields are of  interest
 as they naturally appear in extended supergravities (see, e.g., \cite{Sgrav} and references therein).

The auxiliary-tensor formalism for U(1) self-duality \cite{IZ}-\cite{IZ3}
is based on extending the pure ${F}^{mn}$ system by an auxiliary bispinor field $(V_{\alpha\beta}, \bar V_{\dot\alpha\dot\beta}) \sim V^{mn}$
which is not subject to any extra constraints like the Bianchi identity. In the extended formulation, U(1) self-duality is
equivalent to  manifest U(1) invariance of the nonlinear auxiliary interaction $E(V)$. Solving the algebraic equations for
$V_{\alpha\beta}$ as $V_{\alpha\beta} = V_{\alpha\beta}({F})$,  we regain the self-dual nonlinear Lagrangian $\hat{L}(F)$ in the standard representation.
The main advantage of the formulation with auxiliary tensor fields is that it reduces the
construction of the most general
duality-invariant action for ${F}^{mn}$ to listing all  U(1) invariant auxiliary interactions $E(V)$.
 Later on, this approach was generalized to the U($N$) case \cite{IZ4} and to self-dual systems of nonlinear ${\cal N}=1$ and ${\cal N}=2$
 electrodynamics \cite{Ku,ILZ,IZ5}.

The natural next step is to include coset fields into the formalism of auxiliary (super)fields, first at the purely bosonic level and then,
taking this as a prerequisite, to pass to the corresponding  superfield extensions. In this paper we address the first part of this program.
Namely, we generalize the formalism of auxiliary tensor fields to Sp(2,$\,\mathbb{R}$) duality invariant theories.

In Section 2 we start
with a review of the Gibbons-Rasheed construction of the axion-dilaton couplings in nonlinear
electrodynamics. In Section 3 we introduce the auxiliary bispinor
complex field $\hat{V}_{\alpha\beta}$ which transforms covariantly under the
nonlinear realization (NR) of Sp(2,$\,\mathbb{R}$). By construction,
the basic Lagrangian obeys the Gaillard-Zumino representation
and includes the Sp(2,$\,\mathbb{R}$) invariant nonlinear interaction
of the auxiliary fields $\hat{V}_{\alpha\beta}$. Solving the auxiliary-field equation in terms
of the electromagnetic field $(F_{\alpha\beta}, F_{\dot\alpha\dot\beta})$ we obtain the general self-dual  Sp(2,$\,\mathbb{R}$) Lagrangian
in the standard representation. A more convenient construction of the
auxiliary-field representation is based on a Legendre transformation.
We also consider Sp(2,$\,\mathbb{R}$) self-dual models with higher
derivatives, employing covariant derivatives of the NR auxiliary fields.
Section 4 describes an alternative formalism, which starts from auxiliary bispinor
fields $V$ transforming under  a  {\it linear\/} realization (LR) of Sp(2,$\,\mathbb{R}$).
The auxiliary interaction $E$ in this formulation satisfies nonlinear
constraints. The Legendre transformation allows one to linearize and solve these constraints, connecting $E$ with the Sp(2,$\,\mathbb{R}$) invariant
interaction of auxiliary scalar fields.
The two different choices of the auxiliary bispinor fields lead to equivalent $F$ representations of self-dual
Lagrangians.

\setcounter{equation}0
\section{Axion-dilaton coupling and nonlinear realization of Sp(2,$\,\mathbb{R}$)}

The infinitesimal transformation of the group Sp(2,$\,\mathbb{R}$) $\simeq$
SL(2,$\,\mathbb{R}$) (i.e. an element of the algebra sp(2,$\,\mathbb{R}$))
can be parametrized by the $2\times 2$ matrix
\bea
&&{\cal B}
=\left(\begin{array}{cc}a&b\\
c& -a\end{array}\right),\lb{A1}
\eea
where $a, b$ and $ c$ are real numbers. The  nonlinear realization of Sp(2,$\,\mathbb{R}$) is arranged
as the transformations of the scalar field $S=S_1+iS_2$ \cite{GR2}
\bea
&&\delta S=b+2a S-cS^2,\label{Stransf}\\
&&\delta S_1=b+2aS_1-c(S_1^2-S_2^2), \quad \delta S_2=2(a-cS_1)S_2\,,\lb{A2}
\eea
where the real  scalar fields $S_1$ and $S_2$ are connected with the axion $A$ and dilaton,
$\phi$
\be
S_1=2A,\quad S_2=e^{-2\phi}.
\ee

The invariant K\"{a}hler $\sigma$ model Lagrangian contains the K\"{a}hler metric $g_{S\bar{S}} = -\frac{1}{(S - \bar S)^2}\,$,
\bea
L(S)=g_{S\bar{S}}\partial^m\bar{S}\partial_mS = -\frac{\partial^m\bar{S}\partial_mS}{(S-\bar{S})^2}
=  \frac{(\partial_m S_1)^2 + (\partial_m S_2)^2}{4\,S_2^2} =(e^{2\phi}\partial_m A)^2+(\partial_m\phi)^2\,. \lb{A3}
\eea
It is convenient to make the rescaling
\be
L[S(A,\phi)]~\rightarrow~\frac1{\xi^2}L[S(\xi A,\xi \phi)]\,,
\ee
where $\xi$ is a coupling constant of dimension $-1$.

The scalar equation of motion
\bea
E^0(S) := \frac{\Delta L_0(S)}{\Delta S} =\frac1{(S-\bar{S})^2}\left[\Box\bar{S}
+\frac{2\partial^m\bar{S}\partial_m\bar{S}}{(S-\bar{S})} \right]=0\,, \lb{eomS}
\eea
where $\Delta/\Delta S$ is the Euler-Lagrange derivative, transforms covariantly under the transformations \p{Stransf}:
 \bea
 &&\delta E^0(S)=-2(a-cS)\,E^0(S).
\eea

For the electromagnetic field strengths we will use both the bispinor and tensor representations
\bea
&&F_\alpha^\beta(A)\equiv
\sfrac14\partial_{\alpha\dot\beta}A^{\dot\beta\beta}
-\sfrac14\partial^{\dot\beta\beta}A_{\alpha\dot\beta}=
\sfrac18(\sigma^m\bar\sigma^n-\sigma^n\bar\sigma^m)_\alpha^\beta {F}_{mn},\lb{A4}\\
&&\varphi :=F^{\alpha\beta}F_{\alpha\beta}=t+iz,\quad
\bar\varphi=\bar{F}^2=t-iz,\nn
&& t :=\sfrac14{F}^{mn}{F}_{mn},\quad
z :=\sfrac14{F}^{mn}\tilde{F}_{mn}.\lb{A5}
\eea
The general U(1) self-dual Lagrangian $\hat{L}(\varphi,\bar\varphi)=L^\prime(t,z)$ satisfies
the nonlinear differential condition \cite{GZ,GR}
\bea
\mbox{Im}\, [\varphi - 4\varphi(\hat{L}_\varphi)^2] =0\,.\lb{A6}
\eea

In this notation, the Gibbons-Rasheed (GR) Lagrangian \p{Sp2Rlagr} describing the interaction of scalar fields with the electromagnetic
field in the nonlinear electrodynamics \cite{GR2} is rewritten as
\bea
L^{\rm sd}(S,F) &=& L(S)-\sfrac{i}2S_1\varphi+\sfrac{i}2S_1\bar\varphi
+\hat{L}(\hat\varphi, \hat{\bar\varphi})\nn
&=&\,L(S)+S_1z+L^\prime(S_2t,S_2z) := L(S) + \tilde{L}^{\rm sd}(S, F)\,.\lb{A7}
\eea
Here,  $\hat{L}(\hat{\varphi}, \hat{\bar\varphi})$ is the same U(1) self-dual Lagrangian as before, but with the rescaled arguments,
\bea
\hat\varphi=S_2\varphi\,,\qquad
\hat{\bar\varphi}=S_2\bar\varphi\,. \lb{A8}
\eea
Evidently,  $\hat{L}(\hat\varphi,\hat{\bar\varphi})$ satisfies the same condition as \p{A6}:
\bea
\mbox{Im}\,[\hat\varphi -
4\,\hat\varphi(\hat{L}_{\hat\varphi})^2]=0\,.\lb{A9}
\eea
In what follows, we will need the explicit expression for the part of \p{A7} linear in $\varphi$ and $\bar\varphi$:
\bea
L^{\rm sd}_2 (S, F) = \tilde{L}^{\rm sd}_2 (S, F) = -\sfrac{i}2 (\bar S\varphi - S\bar\varphi) = S_1 z - S_2t\,, \lb{L2SF}
\eea
where we took into account that
\bea
\hat{L}(\varphi, \bar\varphi) = -\sfrac12(\varphi +\bar\varphi) + \hat{L}^{int}(\varphi, \bar\varphi)\,. \lb{L2Lint}
\eea

Now we consider the Bianchi identity for $F_{\alpha\beta}, \bar F_{\dot\alpha\dot\beta}$
\be
B_{\alpha\dot\alpha}=\partial_\alpha^{\dot\beta}
\bar{F}_{\dot\alpha\dot\beta} -\partial^\beta_{\dot\alpha}
F_{\alpha\beta}= 0 \lb{A9b}
\ee
together with the nonlinear $F$-equation of motion
\bea
&&
E_{\alpha\dot\alpha}(S,F)=\partial_\alpha^{\dot\beta}
\bar{P}_{\dot\alpha\dot\beta} -\partial^\beta_{\dot\alpha}
P_{\alpha\beta}= 0\,, \lb{A10}
\eea
where the dual bispinor field $P_{\alpha\beta}$ and its conjugate $\bar{P}_{\dot\alpha\dot\beta}$ are defined by
\bea
&&P_{\alpha\beta}(S,F)=i\frac{\partial L^{\rm sd}}{\partial
F^{\alpha\beta}}
= (S_1+2iS_2\hat{L}_{\hat\varphi})\,F_{\alpha\beta} =:
\sfrac18(\sigma^m\bar\sigma^n-\sigma^n\bar\sigma^m)_{\alpha\beta} {G}_{mn}\,.\lb{A10b}
\eea
The whole dependence of the set of equations \p{A9b}, \p{A10b} on the Sp(2,$\,\mathbb{R}$)  scalar fields $S_1, S_2$
is hidden in the tensors $P_{\alpha\beta}$ and $\bar{P}_{\dot\alpha\dot\beta}\,$.

The self-duality condition \p{A9} guarantees the self-consistency of
the linear Sp(2,$\,\mathbb{R}$)  transformations mixing $F_{\alpha\beta}$ with $P_{\alpha\beta}$
\bea
&&\delta P_{\alpha\beta}=aP_{\alpha\beta}+bF_{\alpha\beta}\,,\lb{A12}\\
&& \delta F_{\alpha\beta}=cP_{\alpha\beta}-aF_{\alpha\beta}
=-(a - cS_1 -2icS_2\hat{L}_{\hat\varphi})\,F_{\alpha\beta}\,.\lb{A13}
\eea
The pair of equations \p{A9b} and \p{A10} is covariant under these Sp(2,$\,\mathbb{R}$) duality transformation.
Eqs. \p{A12}, \p{A13} mean  that the bispinors  $F_{\alpha\beta}$ and $P_{\alpha\beta}$ form a linear  Sp(2,$\,\mathbb{R}$)
doublet, so it is natural to use the notation $R^a :=(P,F)$ and rewrite \p{A12}, \p{A13} as $\delta R^a={\cal B}^a_bR^b$, where
the matrix ${\cal B}^a_b$ was defined in \p{A1}.

For further use we will also introduce the modified NR field strength and dual field strength
\bea
\hat{P}_{\alpha\beta}=i\frac{\partial\hat{L}}{\partial\hat{F}^{\alpha\beta}}
=2i\hat{F}_{\alpha\beta}\hat{L}_{\hat\varphi},\quad
\hat{F}_{\alpha\beta}=\sqrt{S_2}F_{\alpha\beta}\,, \lb{A14}
\eea
in terms of which the  self-duality condition \p{A9} takes the conventional form \cite{GZ}-\cite{KT}:
\be
\mbox{Im}(\hat{P}^2+\hat{F}^2)=0\,.\lb{A17}
\ee

The modified quantities \p{A14} transform nonlinearly  under Sp(2,$\,\mathbb{R}$).  The relation between the NR representation \p{A14}
and the linearly transforming
fields $P_{\alpha\beta}$ and $F_{\alpha\beta}$ can be written in the matrix form as
\bea
\hat{R} =\left(\begin{array}{c}\hat{P}\\
\hat{F}\end{array}\right)=g\left(\begin{array}{c}P\\
F\end{array}\right),\quad R=\left(\begin{array}{c}P\\
F\end{array}\right)=g^{-1}\left(\begin{array}{c}\hat{P}\\
\hat{F}\end{array}\right).
\eea
These relations contain the real coset matrix $g$ and its inverse $g^{-1}(S)$:
\bea
g(S)=\left(\begin{array}{cc}g_{1}^1&g_{2}^1\\
g_{1}^2& g_{2}^2\end{array}\right)=\frac1{\sqrt{S_2}}\left(\begin{array}{cc}1&-S_1\\
0& S_2\end{array}\right),\quad g^{-1}(S)=\frac1{\sqrt{S_2}}\left(\begin{array}{cc}S_2&S_1\\
0& 1\end{array}\right).\lb{gcoset}
\eea
Transformations of the coset matrix have the following form:
\bea
&&\delta g=\Theta g-g{\cal B},\lb{gtrans}\\
&&\Theta=(\delta g)g^{-1}+g{\cal B}g^{-1}=\left(\begin{array}{cc}0&-\rho\\
\rho& 0\end{array}\right)=-i\rho\tau_2\,,
\eea
where $\rho= cS_2$ is the induced  parameter of the nonlinear realization, and
$\tau_2$ is the Pauli matrix. Thus the NR fields transform
covariantly under the nonlinear realization of Sp(2,$\,\mathbb{R}$)
\bea
\delta\hat{P}_{\alpha\beta}=-\rho\hat{F}_{\alpha\beta},\quad
\delta\hat{F}_{\alpha\beta}=\rho\hat{P}_{\alpha\beta}\,. \lb{A16}
\eea
The same transformations can be directly derived from the definition \p{A14} with taking into account
the compatibility constraint \p{A9}. For what follows, it is useful to explicitly write how $\hat{P}_{\alpha\beta}$ is expressed
through $P_{\alpha\beta}$ and $F_{\alpha\beta}$
\bea
\hat{P}_{\alpha\beta} = \frac{1}{\sqrt{S_2}}\left(P_{\alpha\beta} - {S_1}F_{\alpha\beta}\right)
\eea
(the connection between $\hat{F}_{\alpha\beta}$ and $F_{\alpha\beta}$ was already given in
\p{A14}).

It is easy to construct, out of the coset fields $S_1, S_2$, the $2\times 2$ matrix $M_{ab}$ supporting a linear realization (LR) of
Sp(2,$\,\mathbb{R}$):
\bea
&&M(S)=g^Tg=\frac1{S_2}\left(\begin{array}{cc}1&-S_1\\
-S_1& S^2_1+S^2_2\end{array}\right),\quad\mbox{det}M=1\,,\lb{Mmatr}\\
&&\delta M=-{\cal B}^TM-M{\cal B}\,.\nonumber
\eea
One can alternatively write the coset Lagrangian \p{A3} through this matrix
\bea
L(S) = -\sfrac14\,\partial_m M_{ab}\partial^mM_{cd}\varepsilon^{ac}\varepsilon^{bd} \sim {\rm Tr}\,\partial_m M\partial^m M^{-1}\,.
\eea
The matrix of K\"ahler complex structure $J^a_c$ also admits a simple expression through $M$
\be
J^a_c(S)=M_{cb}(S)\varepsilon^{ba},\quad J^a_cJ^c_r=-\delta^a_r.\lb{Jmatr}
\ee

The standard U(1) duality relations are reproduced from the Sp(2,$\,\mathbb{R}$) ones given above in the non-singular
limit
\bea
S_1=0\,, \quad  S_2=1\,. \lb{plane}
\eea
In this limit, $g^a_b \rightarrow  \delta^a_b\,, \; M_{ab} \rightarrow \delta_{ab}$ and Sp(2,$\,\mathbb{R}$) is reduced to its
$O(2)\sim U(1)$ subgroup with $a=0, b= -c$, which preserves \p{plane} and is just the standard U(1) duality group.

\setcounter{equation}0
\section{Nonlinear auxiliary fields in Sp(2,$\,\mathbb{R}$) duality}
\subsection{Sp(2,$\,\mathbb{R}$) duality as invariance of the auxiliary interaction}

We introduce the following complex combinations of the NR bispinor fields $\hat{F}$ and $\hat{P}$ defined in \p{A14}:
\bea
&&\hat{V}_{\alpha\beta}=\sfrac12(\hat{F}+i\hat{P})_{\alpha\beta}\,,\quad \delta\hat{V}_{\alpha\beta}=
-i\rho\hat{V}_{\alpha\beta}\,,\nn
&&\hat{\bar{V}}_{\dot\alpha\dot\beta}=\sfrac12(\hat{\bar{F}}-i\hat{\bar{P}})_{\dot\alpha\dot\beta}\,,\quad
\delta\hat{\bar{V}}_{\dot\alpha\dot\beta}=
i\rho\hat{\bar{V}}_{\dot\alpha\dot\beta}\,.\lb{rhohatV}
\eea
We will treat these fields as independent auxiliary tensor variables of the NR
{\it representation\/} of the tensor formulation of the Sp(2,$\,\mathbb{R}$) self-duality and {\it postulate\/} for
$\hat{F}_{\alpha\beta}, \hat{P}_{\alpha\beta}$ just the NR transformation properties \p{A16}. The standard expression of $\hat{P}$
through the original fields $F_{\alpha\beta}, S_1, S_2$ as given in \p{A14}
will arise after eliminating $\hat{V}_{\alpha\beta}, \hat{\bar{V}}_{\dot\alpha\dot\beta}$ from the appropriate extended action by
their equations of motion. In this extended formulation, we will also use the basic transformation of the NR
electromagnetic field
\bea
&&\delta \hat{F}_{\alpha\beta}=i\rho(\hat{F}-2\hat{V})_{\alpha\beta}, \quad \rightarrow \quad \delta(\hat{F}-\hat{V})_{\alpha\beta}=
i\rho(\hat{F}-\hat{V})_{\alpha\beta}\,,\nonumber
\eea
which is just the result of substituting $P_{\alpha\beta} = i(\hat{F}_{\alpha\beta} - 2 V_{\alpha\beta})$ from the definition \p{rhohatV} into
\p{A16}. The scalar combinations of the auxiliary fields and their transformation laws are given by
\bea
&&\hat\nu=\hat{V}^{\alpha\beta}\hat{V}_{\alpha\beta},\quad \delta \hat\nu=-2i\rho\hat\nu,\nn
&&\hat{\bar\nu}=\hat{\bar{V}}^{\dot\alpha\dot\beta}\hat{\bar{V}}_{\dot\alpha\dot\beta},\quad
\delta \hat{\bar\nu}=2i\rho\hat{\bar\nu},\\
&&\hat{a}=\hat\nu\hat{\bar\nu},\qquad \delta\hat{a}=0\,.\lb{hata}
\eea

By analogy with the auxiliary tensor field formulation  of the U(1) duality \cite{IZ}-\cite{IZ3},
in constructing the extended action we start by defining  the bilinear in $F$ and $\hat{V}$ part of the interaction with scalar fields
\bea
&&{\cal L}_2(S,F,\hat{V})=\sfrac12(S_2-iS_1)F^2+\hat{V}^2- 2\,\sqrt{S_2}(\hat{V}\cdot F)
+\mbox{c.c.}\, .\lb{calL2}
\eea
By construction, the $F$ derivative of this Lagrangian,
\bea
\frac{\partial{\cal L}_2}{\partial F^{\alpha\beta}} =: -iP_{\alpha\beta}(S,F,\hat{V})
=(S_2-iS_1)F_{\alpha\beta}- 2\,\sqrt{S_2}\hat{V}_{\alpha\beta}
=-i(\sqrt{S_2}\hat{P}_{\alpha\beta}+S_1F_{\alpha\beta})
\eea
(where we once more used \p{rhohatV}), together with the field $F_{\alpha\beta}$, transform linearly, according
to the transformation laws \p{A12} and \p{A13}. This implies, in particular, that the modified equation of motion
for the electromagnetic field calculated from ${\cal L}_2$ transform through the Bianchi identity \p{A9b} for $F_{\alpha\beta},
\bar{F}_{\dot\alpha\dot\beta}$ and vice versa, thus exhibiting the Sp(2,$\,\mathbb{R}$) self-duality at this level.
The complex bispinor field
\bea
V_{\alpha\beta}(S,F,\hat{V})=\sfrac12(F+iP)_{\alpha\beta}=
\sfrac12[(1-S_2+iS_1)F_{\alpha\beta}+2\sqrt{S_2}\hat{V}_{\alpha\beta}]\lb{VFhatV}
\eea
also transforms linearly
\bea
&&\delta V_{\alpha\beta}=(a-ic)V_{\alpha\beta}+\sfrac12(ib+ic-2a)F_{\alpha\beta}\,,
\lb{VLR1}\\
&&\delta F_{\alpha\beta}=-2icV_{\alpha\beta}+(ic-a)F_{\alpha\beta}\,.\lb{VLR2}
\eea
Varying ${\cal L}_2$ with respect to $\hat{V}_{\alpha\beta}$
we obtain the  Sp(2,$\,\mathbb{R}$)-covariant auxiliary equation
\bea
\hat{V}_{\alpha\beta}=\sqrt{S_2}F_{\alpha\beta}=\hat{F}_{\alpha\beta}\,.
\eea
Substituting this solution back into ${\cal L}_2\,$, we reproduce the bilinear interaction
of the electromagnetic field with scalars $L^{\rm sd}_2(S, F)$ (eq. \p{L2SF}).

The  bilinear Lagrangian \p{calL2} admits the Gaillard-Zumino-type representation
\bea
&&{\cal L}_2=-\sfrac{i}2(F\cdot P)+I_2+\mbox{c.c.},\\
&&I_2=[\hat{V}^2-\sqrt{S_2}(F\cdot \hat{V})]=-\sfrac14\,R^bR^c\,M_{bc}(S),\\
&&R^1=P,\quad R^2=F\,,
\eea
where $I_2$ is the complex Sp(2,$\,\mathbb{R}$) invariant, and $M_{bc}(S)$ is the LR matrix \p{Mmatr}.
This representation allows us to easily find the  Sp(2,$\,\mathbb{R}$) variation of the bilinear  Lagrangian,
\bea
\delta{\cal L}_2=\sfrac{i}2c(\bar{P}^2-P^2)+\sfrac{i}2b(\bar{F}^2-F^2)\,.
\eea

Now we are prepared to write the total nonlinear Lagrangian in the $(S,F,\hat{V})$ representation. It is a sum of ${\cal L}_2$ and
the Sp(2,$\,\mathbb{R}$) invariant terms
\bea
&&{\cal L}(S,F,\hat{V})=L(S)+{\cal L}_2(S,F,\hat{V})+{\cal E}(\hat{a})\,,\lb{calLtot}\\
&&\frac{\partial{\cal L}}{\partial F^{\alpha\beta}}=
\frac{\partial{\cal L}_2}{\partial F^{\alpha\beta}}
=-iP_{\alpha\beta}(S,F,\hat{V})\,,\nonumber
\eea
where $\hat{a}$ is the invariant quartic auxiliary variable defined in \p{hata}. Since the equations of motion
for the electromagnetic field are not modified as compared to  the ${\cal L}_2$ case, they still exhibit, together with the Bianchi identity,
the Sp(2,$\,\mathbb{R}$) covariance. The equation for the auxiliary field
\bea
(\hat{F}-\hat{V})_{\alpha\beta}=\hat{V}_{\alpha\beta}\hat{\bar\nu}
\frac{d{\cal E}}{d\hat{a}}
\lb{auxeq2}
\eea
is also manifestly Sp(2,$\,\mathbb{R}$) covariant. It is analogous to the twisted self-duality constraints considered in \cite{BN,CKR}.
Note that the equation of motion for the coset fields \p{eomS} is modified by a non-zero source depending on
the fields $\hat{V}, F$. It is easy to show that the Sp(2,$\,\mathbb{R}$) covariance of \p{eomS} is not affected by this modification, like
in the original GR framework.

It is instructive to rewrite \p{calLtot} in the more  detailed form
\bea
{\cal L}(S,F,\hat{V})=L(S) -\sfrac{i}2S_1(\varphi - \bar\varphi) + \sfrac12(\hat\varphi + \hat{\bar\varphi}) +
(\hat{\nu} + \hat{\bar\nu}) - 2[(\hat{V}\cdot \hat F) + (\hat{\bar V}\cdot \hat{\bar F})]  + {\cal E}(\hat a)\,.\lb{calLtot1}
\eea
The sum of the last four terms in \p{calLtot1} precisely coincides with the extended U(1) self-dual Lagrangian of nonlinear electrodynamics of
refs. \cite{IZ} - \cite{IZ3}, up to the rescaling $F_{\alpha\beta} \rightarrow \hat{F}_{\alpha\beta} = \sqrt{S_2}F_{\alpha\beta}$. Hence,
by reasoning of these papers, it should yield the most general self-dual Lagrangian $\hat{L}(\hat\varphi, \bar\varphi)$
upon eliminating  the auxiliary fields $\hat{V}_{\alpha\beta}, \hat{\bar V}_{\dot\alpha\dot\beta}$ by their equations of motion. We conclude that
\p{calLtot}, \p{calLtot1} indeed yields the auxiliary bispinor field extension of the general Sp(2,$\,\mathbb{R}$) self-dual
GR Lagrangian \p{A7}. Let us point out that the whole information about the given Sp(2,$\,\mathbb{R}$) self-dual system is encoded
in the  Sp(2,$\,\mathbb{R}$) invariant function ${\cal E}(\hat{a})$ which is not subject to any constraints. Given some bispinor
field representation of the standard U(1) self-dual action, we can promote it to that defining an Sp(2,$\,\mathbb{R}$) self-dual system just
according to the recipe \p{calLtot1}.

The auxiliary equation \p{auxeq2} is solved by
\bea
\hat{V}_{\alpha\beta}=\hat{F}_{\alpha\beta}\hat{G}(\hat\varphi,
\hat{\bar\varphi}), \quad \hat{G}= \frac{1}{1 + \hat\nu \frac{d{\cal E}}{d \hat{a}}}= \frac12-\frac{\partial\hat{L}}{\partial\hat\varphi}\,.\lb{hatG}
\eea
By analogy with the U(1) case \cite{IZ3} we can use the perturbative
expansion for ${\cal E}(\hat{a})$
\be
{\cal E}(\hat{a})=e_1\hat{a}+\sfrac12e_2\hat{a}^2+\ldots\,,
\ee
where $e_1, e_2,\ldots$ are some constant coefficients. The corresponding perturbative
solution for $\hat{L}$ reads
\bea
&&\hat{L}(\hat\varphi,\hat{\bar\varphi})=-\sfrac12(\hat\varphi+\hat{\bar\varphi})
+e_1\hat\varphi\hat{\bar\varphi}
-e_1^2(\hat\varphi^2\hat{\bar\varphi}+\hat\varphi\hat{\bar\varphi}^2)
+e_1^3(\hat\varphi^3\hat{\bar\varphi}+\hat\varphi\hat{\bar\varphi}^3)
\nn
&& +(4e_1^3+\sfrac12e_2)\hat\varphi^2\hat{\bar\varphi}^2
-e_1^4(\hat\varphi^4\hat{\bar\varphi}+\hat\varphi\hat{\bar\varphi}^4)
-(10e_1^4+2e_1e_2)(\hat\varphi^3\hat{\bar\varphi}^2
+\hat\varphi^2\hat{\bar\varphi}^3)+\ldots.\lb{expanL}
\eea

Like in the U(1) duality case, it is useful to define the intermediate (on-shell) representation for the
self-dual Lagrangian by expressing (formally) the field $\hat{F}$ in terms of $\hat{V}$ from the algebraic
equation \p{auxeq2}:
\bea
&&F_{\alpha\beta}\rightarrow F_{\alpha\beta}(\hat{V})=
\frac1{\sqrt{S_2}}(1+\hat{\bar\nu}\frac{d{\cal E}}{d\hat{a}})\hat{V}_{\alpha\beta}\,,
\lb{auxhatV}\\
&&\hat{P}_{\alpha\beta}=i(\hat{F}-2\hat{V})_{\alpha\beta}\rightarrow
\hat{P}_{\alpha\beta}(\hat{V})=i[\hat{F}(\hat{V})-2\hat{V}]_{\alpha\beta}\,.\nonumber
\eea
This ``on-shell'' representation preserves the Sp(2,$\,\mathbb{R}$) covariance
\bea
\delta[\hat{F}(\hat{V})-\hat{V}]_{\alpha\beta}=
i\rho[\hat{F}(\hat{V})-\hat{V}]_{\alpha\beta}\,.
\eea
The substitution \p{auxhatV} gives
\bea
&&{\cal L}_2(S,F,\hat{V})\rightarrow
\sfrac12(S_2-iS_1)[F(\hat{V})\cdot F(\hat{V})]+\hat{V}^2- 2\,\sqrt{S_2}[\hat{V}\cdot F(\hat{V})]
+\mbox{c.c.}\nn
&&=-\frac{i}2[P(\hat{V})\cdot F(\hat{V})] -\hat{a}\frac{d{\cal E}}{d\hat{a}}
+\mbox{c.c.}\,.
\eea
The same transform applied to the total Lagrangian \p{calLtot} preserves the GZ form of the latter
\bea
&&{\cal L}(S,F,\hat{V})\rightarrow\tilde{\cal L}(S,\hat{V})=L(S)
+\sfrac{i}2[\bar{P}(\hat{V})\cdot \bar{F}(\hat{V}) - P(\hat{V})\cdot F(\hat{V})]
-2\hat{a}\frac{d{\cal E}}{d\hat{a}}+{\cal E}(\hat{a}).\lb{tildeLhatV}
\eea
Substituting here the solution $\hat{V}(\hat{F})$ \p{hatG}, one recovers the
 $F$ representation of the Lagrangian, i.e. \p{A7}.

Being applied to  eq. \p{VFhatV}, the change \p{auxhatV} yields
\bea
V_{\alpha\beta}(S,F,\hat{V}) &\rightarrow& V_{\alpha\beta}(S,\hat{V})
=\sfrac12[(1-S_2+iS_1)F_{\alpha\beta}(\hat{V})+2\sqrt{S_2}\hat{V}_{\alpha\beta}]\nn
&=& \left[\frac1{2\sqrt{S_2}}
(1+\hat{\bar\nu}\frac{d{\cal E}}{d\hat{a}})(1-S_2+iS_1)+\sqrt{S_2}\right]\hat{V}_{\alpha\beta}\,.
\lb{VhatV}
\eea
This establishes the relation between the LR and NR auxiliary fields (on the shell of the auxiliary equation \p{auxeq2}), which
involves the scalar coset fields and the invariant interaction ${\cal E}$.
The corresponding ``on-shell'' relation between the scalar combinations of the auxiliary
fields reads
\bea
&&\nu(S,\hat{V})=\left[\frac1{2\sqrt{S_2}}
(1+\hat{\bar\nu}\frac{d{\cal E}}{d\hat{a}})(1-S_2+iS_1)+\sqrt{S_2}\right]^2\hat\nu. \lb{nuhatnu}
\eea

\subsection{Legendre transformation for the nonlinear auxiliary fields}

The Legendre transformation for the auxiliary field formulation of the
U(1) self-dual electrodynamics was discussed in \cite{IZ2,IZ3}. This
transformation simplifies solving the auxiliary self-duality equation, which is the central
step in deriving the conventional self-dual Lagrangian from the extended one.

To generalize this to the theory with the Sp(2,$\,\mathbb{R}$) scalars, we introduce some new NR covariant scalar auxiliary
fields $\hat\mu$ and $\hat{\bar\mu}$ with the transformation law
\be
 \delta \hat\mu=2i\rho\hat\mu,\quad \delta
\hat{\bar\mu}=-2i\rho\hat{\bar\mu} \lb{hatmutrans}
\ee
and define the following generalized  Lagrangian simultaneously involving  two types of the auxiliary fields:
\bea
{\cal L}(S,F,\hat{V},\hat\mu) &=& L(S)-\sfrac{i}2S_1S^{-1}_2\hat\varphi
+\sfrac{i}2S_1S^{-1}_2\hat{\bar\varphi} +\hat\nu+ \hat{\bar\nu}-
2\,[(\hat{V}\cdot\hat{F})+(\hat{\bar{V}}\cdot\hat{\bar{F}})]
+\sfrac12(\hat\varphi+\hat{\bar\varphi})\nn
&&+\,\hat\nu\hat\mu+\hat{\bar\nu}\hat{\bar\mu}+I(\hat{b})\,, \lb{LSmu}
\eea
where
$I(\hat{b})$ is a function of the new invariant auxiliary variable
$\hat{b}=|\hat\mu|^2$. This representation guarantees the Sp(2,$\,\mathbb{R}$)
covariance of the electromagnetic-scalar equations. Using the
$\hat{V}$ equation of motion,
\bea
\hat{V}_{\alpha\beta}= \frac1{1+\hat\mu}\,\hat{F}_{\alpha\beta}\,,
\eea
we can eliminate the variable $\hat{V}_{\alpha\beta}$ from \p{LSmu}:
\bea
&&\tilde{\cal L}(S,\varphi,\hat\mu)=L(S) + \left[-\frac{i}2S_1\varphi
+\frac{S_2\varphi(\hat\mu-1)}{2(1+\hat\mu)} +\mbox{c.c.}\right]
+I(\hat{b})\nn
&&=L(S) + \sfrac{i}2\left[(\bar{F}\cdot\bar{P})-(F\cdot P)\right]
+I(\hat{b})\,. \lb{Lvm}
\eea
In this case we obtain the simple expressions for the dual fields
\bea
P_{\alpha\beta}= i\frac{\partial \tilde{\cal L}}{\partial F^{\alpha\beta}} = \left[S_1+\frac{iS_2(\hat\mu-1)}{(1+\hat\mu)}\right]F_{\alpha\beta}
=\sqrt{S_2}\hat{P}_{\alpha\beta}
+\frac{S_1}{\sqrt{S_2}}\hat{F}_{\alpha\beta}\,,\lb{Phatmu}
\eea
whence
\bea
\hat{P}_{\alpha\beta}=\frac{i(\hat\mu-1)}{1+\hat\mu}\,\hat{F}_{\alpha\beta}\,.
\eea
These explicit expressions provide the correct transformation laws for the relevant quantities.

In fact, the representation \p{LSmu} just defines the Legendre transformation of the
Lagrangian \p{calLtot1}. Indeed, varying \p{LSmu} with respect to the auxiliary field $\hat\mu$ yields
\bea
\hat\nu = -\frac{\partial I(\hat{b})}{\partial \hat{\mu}}\,.
\eea
Using this equation,  $\hat{\mu}$ and $\hat{b}$ can be expressed in terms of $\hat{\nu}, \hat{\bar\nu}$, assuming that $(\frac{dI}{d\hat{b}})^{-1}$
is not singular at $\hat{b}=0$.
After the elimination of $\mu, \bar\mu$, the Lagrangian \p{LSmu} takes the form of \p{calLtot1} with
\bea
{\cal E}(\hat{a}) = I(\hat{b}) - 2\hat{b}\frac{d{I}}{d\hat{b}}\,.
\eea
Then it is easy to show that
\bea
\hat{\mu} = \frac{\partial {\cal E}(\hat{a})}{\partial \hat{\nu}}\,.
\eea

Varying $\tilde{\cal L}(S,\hat\varphi,\hat\mu)$ \p{Lvm} with respect to $\hat\mu\,$, we arrive at the
equation for the auxiliary scalar variables
\bea
\hat{\varphi}=S_2\varphi=-(\hat{\bar\mu} +2\hat{b}+
\hat{b}\,\hat\mu)\, \frac{dI}{d\hat{b}}\,. \lb{varmu}
\eea
It is analogous to the corresponding equation in the U(1) self-dual theory.
Solving it for $\hat\mu = \hat\mu(\hat\varphi, \hat{\bar\varphi})$, we obtain the equation for determining the
self-dual Lagrangian $\hat{L}(\hat\varphi,\hat{\bar\varphi})$
\be
\frac1{1+\hat\mu}=\frac12-\frac{\partial\hat{L}}{\partial\hat\varphi}\,.
\ee

By analogy with \cite{IZ2,IZ3}, as a notorious example, we can obtain the exact Lagrangian
for the Born-Infeld (BI) theory extended by the Sp(2,$\,\mathbb{R}$) scalar fields, proceeding from the invariant interaction
\be
 I(\hat{b})=\frac{2\hat{b}}{\hat{b}-1}\,.
\ee
In this case, one can find the complete solution to the equation \p{varmu}
\bea
\hat\mu &=& \frac{Q-1-\frac12(\hat\varphi-\hat{\bar\varphi})}
{Q+1+\frac12(\hat\varphi-\hat{\bar\varphi})},\\
Q(\hat\varphi, \hat{\bar\varphi}) &=& \sqrt{1+(\hat\varphi+\hat{\bar\varphi})
+\sfrac14(\hat\varphi-\hat{\bar\varphi})^2}=Q(S_2\varphi, S_2 \bar\varphi)=1-L_{BI}(S_2\varphi, S_2 \bar\varphi).
\eea
Taking $L_{BI}(S_2\varphi, S_2\bar\varphi)$ as $\hat{L}(\hat\varphi, \bar{\hat{\varphi}})$ in eq. \p{A7},
we recover the standard GR coupling of the scalar fields in the BI theory \cite{GR2}. Note that the derivation
of the BI Lagrangian in the $\hat{b}$ representation is much easier than in the original $\hat{a}$ formulation.

All other examples of the U(1) self-dual systems in which eq. \p{varmu} has a closed solution can also be generalized
to the Sp(2,$\,\mathbb{R}$) case. For instance, the invariant auxiliary interaction
\be
 I(\hat{b})=2\ln(1-\hat{b})
\ee
 yields the cubic equation for $\hat\mu(\hat\varphi)$ and gives us the exact expression for
$\hat{L}(\hat\varphi, \hat{\bar\varphi})$, like in the analogous U(1) model considered in \cite{IZ3}.

The scalar coupling in the so called ``simplest interaction model'' \cite{IZ,BN,IZ3} corresponds to the
choice $I_{SI}=-2\hat{b}\,$.

\subsection{Sp(2,$\,\mathbb{R}$) duality and higher derivatives}

The self-dual theories with higher derivatives in the standard setting were analyzed in
\cite{BN,CKO}. The formulation through auxiliary tensor fields for such theories was worked out in \cite{IZ3}.
As was shown there, any U(1) self-dual theory
with higher derivatives is generated by the appropriate U(1) invariant auxiliary interaction involving
space-time derivatives of the auxiliary bispinor fields .

The transformation parameter $\rho=cS_2$ in the NR representation \p{rhohatV}
depends on the scalar field, so in the case under consideration we need to properly Sp(2,$\,\mathbb{R}$) covariantize
the space-time derivatives of the auxiliary fields involved. This can be done following the standard routine of the
nonlinear realizations \cite{Nonl}.

First, we construct the $2\times2$ matrix Cartan 1-forms pertinent to the nonlinear realization of Sp(2,$\,\mathbb{R}$) in the
symmetric coset space Sp(2,$\,\mathbb{R}$)$/$U(1) we deal with:
\bea
&&dg g^{-1}=
\Gamma+D,\lb{B17}\\
&&\Gamma=dx^m\Gamma_m=\sfrac12dg g^{-1}-\sfrac12g^{-1T}dg^T=
\left(\begin{array}{cc}0&-\zeta\\
\zeta& 0\end{array}\right),\lb{B18}\\
&&
D=dx^mD_m=\sfrac12dg g^{-1}+\sfrac12g^{-1T}dg^T=\left(\begin{array}{cc}p&q\\
q& -p\end{array}\right).\lb{B19}
\eea
Here,  $g$ is the coset matrix \p{gcoset} and $g^{-1},~ g^T$ are the
corresponding inverse and transposed matrices. The 1-form $\Gamma=-i\tau_2dx^m\zeta_m$ contains the induced  connection
$\zeta_m(S)$ which defines the covariant derivatives of fields having the standard transformation properties, i.e. transforming with
the induced U(1) parameter $\rho$, while $D$ specifies the NR-covariant   derivative $D_m(S)$ of the coset
fields \footnote{The Lagrangian \p{A3} is just the square of these covariant derivatives, $L(S) = \frac12 {\rm Tr}\, D_m(S) D^m (S)\,$.}.
These objects have the following transformation rules
\bea
&&\delta \Gamma=d\Theta+[\Theta,\Gamma]\,,\quad \delta D=[\Theta,D].\lb{B20}
\eea

The  connection 1-form reads
\bea
&&\zeta=dx^m\zeta_m = \sfrac12(g_{1}^1)^2dS_1\,, \;\;\zeta_m = \frac1{2 S_2}\partial_mS_1\,,\qquad
\delta\zeta_m =\partial_m\rho\,.\lb{NRconn}
\eea
The explicit expressions for the component 1-forms collected in the matrix \p{B19} are as follows
\bea
&&p=dx^mp_m= dg_{1}^1g_{2}^2 = -\frac1{2S_2}\,dS_2 ,\quad
q=dx^mq_m= -\sfrac12 (g_{1}^1)^2 dS_1 = -\frac1{2S_2}\,dS_1  ,
\\
&&\delta p_m=-2\rho q_m,\quad
\delta q_m=2\rho p_m,\quad \delta(p_m+iq_m)=2i\rho(p_m+iq_m).\lb{NRcov}
\eea

Now we are ready to define the covariant derivatives of the NR auxiliary fields
$\hat{V}$ and $\hat{\bar{V}}$ (we suppress their Lorentz indices)
\bea
&&\nabla_m\hat{V}=(\bar\sigma_m)^{\dot\rho\rho}\nabla_{\rho\dot\rho}\hat{V}
=(\partial_m+i\zeta_m)\hat{V},\quad
\delta \nabla_m\hat{V}=-i\rho \nabla_m\hat{V},\\
&&\nabla_m\hat{\bar{V}}=(\bar\sigma_m)^{\dot\rho\rho}\nabla_{\rho\dot\rho}
\hat{\bar{V}}=(\partial_m-i\zeta_m)\hat{\bar{V}},\quad
\delta \nabla_m\hat{\bar{V}}=i\rho \nabla_m\hat{\bar{V}}.
\eea

The corresponding Sp(2,$\,\mathbb{R}$) invariant auxiliary interaction can be constructed by analogy with the U(1) self-dual theory \cite{IZ3}.
We should add to the standard bilinear interaction ${\cal L}_2(S,F,\hat{V})$, eq.  \p{calL2}, the
general Sp(2,$\,\mathbb{R}$) invariant interaction involving the covariant derivatives
of the scalar coset fields and the NR auxiliary fields:
\bea
{\cal E}^K_{der}(\hat{V}, p_m(S), q_m(S),\nabla_m \hat{V}, \ldots\,,
\nabla_{m_1}\nabla_{m_2}\cdots\nabla_{m_k} \hat{V}\,, \ldots)\,.
\eea

The Sp(2,$\,\mathbb{R}$)-covariant local equations of motion for
the auxiliary fields in this
case contain the Euler-Lagrange derivative
\bea
(\hat{V}-\hat{F})_{\alpha\beta}+\frac12\frac{\Delta\hat{\cal E}^K_{der}}
{\Delta \hat{V}^{\alpha\beta}}=0\,.\lb{dertwisd}
\eea
Solving this equation (e.g., by recursions), we finally obtain the Sp(2,$\,\mathbb{R}$) self-dual Lagrangian in the initial
$(F,S)$ representation.

The auxiliary interaction ${\cal E}^K_{der}$ specifying the Sp(2,$\,\mathbb{R}$) self-dual models with higher derivatives
involves new dimensionful constants, starting from the  coupling constant $c$ of dimension $-2$, as well as additional dimensionless
coupling constants. Examples of interaction with two derivatives are provided by the terms
\bea
\sim \nabla_\beta^{\dot\beta}\hat{V}^{\alpha\beta}\nabla_\alpha^{\dot\xi}
\hat{\bar{V}}_{\dot\beta\dot\xi},\qquad \sim \nabla^m\hat\nu\nabla_m\hat{\bar\nu},\ldots\,.
\eea
Non-standard terms with higher derivatives can be generated by the invariant combinations
of the scalar and auxiliary fields, e.g.,
\bea
R_m=(p_m+iq_m)\hat{V}^2\,.
\eea

Note that terms with higher derivatives now also appear in the scalar $S_1, S_2$ equations.

\setcounter{equation}0
\section{Alternative auxiliary-field formulation of Sp(2,$\,\mathbb{R}$) theory}

In the previous section we started from the renowned GR action \p{Sp2Rlagr}, \p{A7} and picked up the nonlinearly
transforming  bispinor auxiliary fields $\hat{V}_{\alpha\beta}$ so as to construct the natural generalization of the extended
formulation of the U(1) self-dual electrodynamics to the case of the Sp(2,$\,\mathbb{R}$) self-dual systems with the coset scalar
fields. In this section we present an alternative construction which starts just from the extended U(1) formulation and produces
the GR action as an output. Its basic distinguishing feature is that it starts with the {\it linear\/} realization of
Sp(2,$\,\mathbb{R}$) on the set $(V_{\alpha\beta}, F_{\alpha\beta})$.

\subsection{$\lambda$ parametrization of the Sp(2,$\,\mathbb{R}$) coset}

In the alternative construction it will be convenient to use another parametrization of the coset of Sp(2,$\,\mathbb{R}$),
this time by the complex scalar field $\lambda(x)$:
\bea
&&\bar{S}(\lambda)=\frac{i(\lambda-1)}{\lambda+1}\,,\qquad
\lambda (S)=\frac{1-i\bar{S}}{1+i\bar{S}},\lb{Slambda} \\
&&S_1(\lambda)=\frac{i(\lambda-\bar\lambda)}{(1+\lambda)(1 +\bar\lambda)},
\qquad
S_2(\lambda)=\frac{1-\lambda\bar\lambda}
{(1+\lambda)(1+\bar\lambda)},
\eea
and use  the alternative set of the group parameters
\bea
&&\alpha=-a-\sfrac{i}2(b+c),\quad \bar\alpha=-a+\sfrac{i}2(b+c),\quad
\gamma=\sfrac12(c-b),\\
&&a=-\sfrac12(\alpha+\bar\alpha),\quad b=\sfrac{i}2(\alpha-\bar\alpha)-\gamma,
\quad c=\sfrac{i}2(\alpha-\bar\alpha)+\gamma.
\eea
The new coset field has a simple transformation law
\bea
\delta\lambda=\delta\frac{2}{1+i\bar{S}}
=\alpha+2i\gamma\lambda
-\bar\alpha\lambda^2\,,\lb{lambtran}
\eea
which resembles the $\mathbb{CP}_1$ realization of the group SU(2),
the only difference being the sign of the last term in \p{lambtran}.

The scalar Lagrangian \p{A3} can be rewritten in terms of $\lambda$ as
\bea
L(S)= L^\prime(\lambda)=\frac{\partial^m\bar\lambda\partial_m\lambda}
{(1-\lambda\bar\lambda)^2}\,. \lb{lambdaAct}
\eea

The relevant coset element can be represented by the Hermitian matrix
\bea
&&G=\frac{1}{\sqrt{1 - \lambda\bar\lambda}}\left(\begin{array}{cc}1&-\lambda\\
-\bar\lambda&1\end{array}\right),\lb{Gcoset}
\eea
with the  transformation law
\bea
&&\delta G=i\tau_3\tilde\gamma G-G\Lambda=-\Lambda^\dagger G-i\tilde\gamma G\tau_3\,,\quad
\tilde{\gamma}(\lambda) := \gamma + \sfrac{1}{2i}(\alpha\bar\lambda -
\bar\alpha\lambda)\,,\lb{Gtran}
\eea
where $\tilde{\gamma}(\lambda)$ is the corresponding induced parameter, $\tau_3$ is the Pauli matrix and
\bea
\Lambda= \left(\begin{array}{cc}i\gamma&\alpha\\
\bar\alpha&-i\gamma\end{array}\right).
\eea

The previously considered Sp(2,$\,\mathbb{R}$) spinors $R^a = (P, F)$ linearly transforming with the matrix ${\cal B}$
are related to the spinors $T^a$ transforming with the new matrix $\Lambda$ as
\bea
&&T=AR
=A\left(\begin{array}{c}P\\
F\end{array}\right),\quad R=A^{-1}T,\quad\delta T=\Lambda T\,,
\eea
where
\bea
&&\Lambda=A{\cal B}A^{-1}= A{\cal B}A^\dagger\,, \quad \Lambda^\dagger=
A{\cal B}^TA^\dagger\,, \\
&& A=\frac{1}{\sqrt{2}}\left(\begin{array}{cc}-i&1\\
i&1\end{array}\right),\quad A^{-1}=\frac1{\sqrt{2}}\left(\begin{array}{cc}i&-i\\
1&1\end{array}\right)= A^\dagger\,.\nonumber
\eea

The Cartan form in this representation, $dG G^{-1}\,$, contains the corresponding connection and covariant derivatives
 by analogy with \p{B17}:
 \bea
 dG G^{-1} = \frac{1}{2(1 -\lambda\bar\lambda)}\left(\begin{array}{cc}\lambda d\bar\lambda -\bar\lambda d\lambda&-2d\lambda\\
-2d\bar\lambda &  - \lambda d\bar\lambda + \bar\lambda d\lambda\end{array}\right).\lb{Cartanlamb}
 \eea
The off-diagonal elements in  \p{Cartanlamb} are just the covariant differentials $D\lambda =dx^m D_m\lambda$ and $D\bar\lambda =dx^m D_m\bar\lambda\,,$
$\delta D\lambda = 2i\tilde{\gamma}D\lambda$ (the Lagrangian
\p{lambdaAct} is bilinear in the covariant derivatives $D_m\lambda\,, D_m\bar\lambda$), while the diagonal element is the U(1) connection
\bea
{\cal A} = {\cal A}_mdx^m = \frac1{2i}\frac{\bar\lambda d\lambda - \lambda d\bar\lambda}{1 -\lambda\bar\lambda}\,,
\quad \delta {\cal A} = d\tilde{\gamma}\,. \lb{Connect}
\eea
It defines the covariant derivative of some field with the standard transformation law under the nonlinear realization considered
\bea
D_m\psi = (\partial_m - i q{\cal A}_m)\psi\,, \quad \delta \psi = i q\, \tilde{\gamma}\psi\,, \;\; \delta D_m\psi =i q \,\tilde{\gamma}D_m\psi\,,
\eea
where $q$ is the U(1) charge of $\psi\,$.

The  real symmetric LR matrix defined in \p{Mmatr} is related to $G^2$
\bea
&&M(\lambda)=A^{-1}G^2A=g^T(\lambda)g(\lambda),\quad
\delta M(\lambda)=-{\cal B}^TM(\lambda)-M(\lambda){\cal B}\,,
\eea
where $g(\lambda):=g[S(\lambda)]$ is the $\lambda$ representation of the real coset matrix  \p{gcoset}.

Two alternative Sp(2,$\,\mathbb{R}$) cosets are  connected by the intertwining matrix
\bea
&&\mbox{Int}:=GAg^{-1},\quad \delta (\mbox{Int})=i\tilde\gamma\tau_3\, \mbox{Int}+i\mbox{Int}\,\tau_2\rho.\lb{tautran}
\eea

\subsection{Auxiliary fields}

The starting point of the alternative construction of the extended Lagrangian for the Sp(2,$\,\mathbb{R}$) self-duality is just
the U(1) Lagrangian of refs. \cite{IZ} - \cite{IZ3}:
\bea
{\cal L}(F, V)= \sfrac12(\varphi + {\bar\varphi}) +
\nu + {\bar\nu} - 2[(V\cdot F) + ({\bar V}\cdot {\bar F})]  + {\cal E}(a)\,, \quad a = \nu\bar\nu\,. \lb{U1orig}
\eea
It is easy to check that the equations of motion and Bianchi identities for $F_{\alpha\beta}$,
\bea
&&\partial_{\dot\alpha}^\beta (F_{\alpha\beta} - 2V_{\alpha\beta}) + {\rm c.c.} = 0\,, \lb{Eq} \\
&& \partial_{\dot\alpha}^\beta\,F_{\alpha\beta} - {\rm c.c.} = 0\,, \lb{Bian}
\eea
are covariant under the Sp(2,$\,\mathbb{R}$) rotations
\bea
\delta V_{\alpha\beta} = -i\gamma\, V_{\alpha\beta} + \bar{\alpha}\,(F-V)_{\alpha\beta}\,, \quad
\delta (F- V)_{\alpha\beta} = i\gamma\, (F-V)_{\alpha\beta} + \alpha\, V_{\alpha\beta}\,,\lb{FVTrans}
\eea
or
\bea
\delta F_{\alpha\beta} = i\gamma ( F_{\alpha\beta} - 2V_{\alpha\beta}) + \alpha V_{\alpha\beta}
+ \bar\alpha (F_{\alpha\beta} - V_{\alpha\beta} )\,. \lb{Ftrans}
\eea
These linear transformations coincide with the transformations \p{VLR1}, \p{VLR2} in the $S_1, S_2$ basis. At the same time, the algebraic equation
\bea
F_{\alpha\beta} - V_{\alpha\beta} = V_{\alpha\beta}\,{\cal E}_\nu\,, \lb{Alg}
\eea
is evidently not covariant. The question is how to modify eq. \p{Alg}
in order to make it also Sp(2,$\,\mathbb{R}$) covariant.

In the free case, with
${\cal E}(\nu, \bar\nu)=0$, this modification is rather simple:
\be
(F - V)_{\alpha\beta} = 0\;\Rightarrow \; F_{\alpha\beta} - (1 +\lambda) V_{\alpha\beta} = 0\,. \lb{Modfree}
\ee
Using the transformation law of $\lambda$ \p{lambtran}, it is easy to show that
\bea
\delta \left[F_{\alpha\beta} - (1 +\lambda) V_{\alpha\beta}\right] =
(i\gamma -\bar\alpha \lambda)\left[F_{\alpha\beta} - (1 +\lambda) V_{\alpha\beta}\right], \lb{lhsTrans}
\eea
which implies the covariance of \p{Modfree}.

As the next step, the following generalization of \p{Modfree} naturally occurs:
\be
F_{\alpha\beta} - (1 +\lambda) V_{\alpha\beta} = V_{\alpha\beta}\,E_\nu\,, \qquad
E = E(\nu,\bar\nu, \lambda,\bar\lambda)\,. \lb{Alggen}
\ee
The function $E(\nu,\bar\nu, \lambda, \bar\lambda)$ is assumed to become the previous ${E}(\nu, \bar\nu) = {\cal E}(a)$
in the limit $\lambda = 0\,$ yielding the Lagrangian \p{U1orig} with the residual U(1) duality group with the parameter $\gamma\,$.
 Eq. \p{Alggen}, together with the dynamical equation \p{Eq}, can be obtained from the Lagrangian
\be
{\cal L}'(\lambda, F, V) = L'(\lambda) + (1 + \lambda)\nu + (1 +\bar\lambda)\bar\nu- 2[(V\cdot F) + (\bar{V}\cdot \bar{F})]   + \sfrac{1}{2}(\varphi + \bar\varphi) +
E(\nu,\bar\nu,\lambda\bar\lambda)\,. \lb{LagrVF}
\ee
For $E =0$, using the algebraic equation \p{Modfree}, we obtain the following $\lambda$-modified free $(F, \bar F)$ Lagrangian:
\be
L_{2}^{\rm sd}(\lambda, F) = \frac12\,\frac{\lambda -1}{\lambda + 1}\,\varphi
+ \frac12\,\frac{\bar\lambda -1}{\bar\lambda + 1}\,\bar\varphi\,.
\ee
With taking into account the relations \p{Slambda}, it coincides with the bilinear part \p{L2SF} of the GR Lagrangian in the $(S_1, S_2)$ parametrization.

In what follows, it will be convenient to deal with the modified interaction function
\be
\hat{E} = E + \lambda\nu + \bar\lambda\bar\nu\,, \lb{hatE}
\ee
that corresponds to transferring the $\lambda$-dependent terms in the l.h.s. of \p{Alggen} to its r.h.s.
 We will fix the $\lambda$-dependence of the interaction $E$ (or $\hat{E}$) from the requirement of compatibility
of the Sp(2,$\,\mathbb{R}$) variations of the left- and right-hand sides of eq. \p{Alggen}. Using the transformation laws
\p{FVTrans}, we find that,  on the shell of the auxiliary equation \p{Alggen},
\bea
\delta \nu = -2i\gamma\,\nu + 2\bar\alpha\, \nu \hat{E}_\nu\,, \quad
\delta \bar\nu = 2i\gamma\,\bar\nu + 2\alpha\, \bar\nu \hat{E}_{\bar\nu}\,.\lb{nuTrans1}
\eea
Then, taking into account this transformation law together with \p{FVTrans}, \p{lambtran} and, once again, \p{Alggen},
we find the variation of the r.h.s. of \p{Alggen} and compare it with that of the l.h.s., i.e., with \p{lhsTrans}.
We find the following conditions\footnote{Actually, the original conditions are the vanishing of the $\nu$ and $\bar\nu$
derivatives of the equations below, but we assume that the integration constants (which depend only on $\lambda$ and $\bar\lambda$)
can be put equal to zero without loss of generality.} on the function $\hat{E}$
\bea
&&\nu \hat{E}_\nu -\bar\nu \hat{E}_{\bar\nu} - \lambda \hat{E}_\lambda + \bar\lambda \hat{E}_{\bar\lambda} = 0\,,\lb{U1lambda1} \\
&&\nu (\hat{E}_\nu)^2 - \bar\nu + D_{\bar\lambda}\hat{E} = 0\,,\lb{baralpha1} \\
&&\bar\nu (\hat{E}_{\bar\nu})^2 - \nu + D_{\lambda}\hat{E} = 0\,,\lb{alpha1}
\eea
where
\be
D_{\bar\lambda} = \partial_{\bar\lambda} - \lambda^2\partial_{\lambda}\,, \;\;
D_{\lambda} = \partial_{\lambda} - \bar\lambda^2\partial_{\bar\lambda}\,, \quad
[D_{\bar\lambda}, D_{\lambda}] = 2(\lambda \partial_\lambda - \bar\lambda\partial_{\bar\lambda})\,.  \lb{defDlambda}
\ee
Eq. \p{U1lambda1} is just the condition of the U(1) invariance of the generalized function $\hat{E}(\nu,\bar\nu, \lambda,\bar\lambda)$.
The mutually conjugated eqs. \p{baralpha1} and \p{alpha1} are new. One can check that the same system of equations
arises from the requirement that the transformations \p{nuTrans1} have the correct $sp(2, \mathbb{R})$ closure.

Repeatedly using the constraints \p{baralpha1} and \p{alpha1}, one finds that
\be
\delta \hat{E}_\nu = \alpha -\bar\alpha (\hat{E}_\nu)^2\,, \quad \delta \hat{E}_{\bar\nu} = \bar\alpha
- \alpha (\hat{E}_{\bar\nu})^2\,, \lb{tranEnu}
\ee
and
\be
\delta \hat{E} = \alpha\left[\nu + \bar\nu(\hat{E}_{\bar\nu})^2 \right] + \bar\alpha\left[\bar\nu + \nu(\hat{E}_{\nu})^2 \right].\lb{tranE}
\ee
An interesting peculiarity is that the transformation laws of $\hat{E}_\nu$ and $\hat{E}_{\bar\nu}$ exactly mimic those of
$\lambda$ and $\bar\lambda$. We also observe that $\hat{E}$ {\it is not invariant\/} under the coset Sp(2,$\,\mathbb{R}$)$/$U(1) transformations,
while it is still invariant under
their U(1) closure. Surprisingly, we can construct such Sp(2,$\,\mathbb{R}$) invariant
from the two independent U(1) invariants
\be
H(\nu, \bar\nu, \lambda, \bar\lambda) = \hat{E}- (\nu \hat{E}_\nu + \bar\nu \hat{E}_{\bar\nu}) =
E- (\nu E_\nu + \bar\nu E_{\bar\nu})\,, \quad \delta H = 0\,.    \lb{defH}
\ee
A similar  object already appeared in \cite{IZ2}, when performing the Legendre transformation from the variables $\nu, \bar\nu$ to
$\mu, \bar\mu$ (recall also Subsection 3.2).

At this step, we deal with  the bispinor field extended Lagrangian \p{LagrVF}, which is reduced to the extended
Lagrangian \p{U1orig} of the U(1) duality systems
in the limit $\lambda =0$ and exhibits Sp(2,$\,\mathbb{R}$) duality under the constraints \p{U1lambda1} - \p{alpha1} on the interaction function
$\hat{E}(\nu, \bar\nu, \lambda, \bar\lambda)\,$. The question is how to solve these constraints via some unconstrained ``prepotential''
which would be analogous to ${\cal E}(\hat{a})$ of Subsection 3.1. While for the time being we do not know how to achieve this, the problem is radically
simplified in the $\mu$ representation obtained by Legendre transformation of \p{LagrVF}, with the  Sp(2,$\,\mathbb{R}$) invariant interaction
$H$, eq. \p{defH}, instead of $\hat{E}\,$. In this representation, the constraints are {\it linearized\/}.

Let us define
\be
\mu := \hat{E}_\nu\,, \quad \bar{\mu} =  \hat{E}_{\bar\nu}\,.
\ee
In the $\mu$ representation:
\be
\nu = -H_\mu\,, \quad \bar\nu = -H_{\bar\mu}, \quad H = H(\mu, \bar{\mu}, \lambda, \bar\lambda)\,.\lb{defHmu}
\ee
The corresponding Legendre-transformed Lagrangian can be derived from
\bea
{\cal L}'(\lambda, F, V, \mu) = L'(\lambda) + \nu + \bar\nu- 2[(V\cdot F) + (\bar{V}\cdot \bar{F})] + \sfrac{1}{2}(\varphi + \bar\varphi) +
\nu\mu + \bar\nu\bar\mu + H(\mu,\bar\mu,\lambda\bar\lambda)\,. \lb{LagrVF1}
\eea
Varying it with respect to $\mu$, $\bar\mu$ yields just eqs. \p{defHmu}, which express $\mu, \bar\mu$ in terms of $\nu, \bar\nu, \lambda, \bar\lambda\,$,
taking us back to \p{LagrVF}. On the other hand, eliminating $V_{\alpha\beta}$, we obtain an analog of the $(F, \mu)$ Lagrangian \p{Lvm}.

It is easy to show that the Sp(2,$\,\mathbb{R}$) invariance conditions \p{baralpha1}, \p{alpha1} are indeed linearized in this $\mu$-representation
\be
D_\lambda H + (\partial_\mu - \bar{\mu}{}^2\partial_{\bar\mu}) H = 0\,, \quad
D_{\bar\lambda} H  + (\partial_{\bar\mu} - {\mu}^2\partial_{\mu}) H = 0\,, \lb{alpha2}
\ee
while \p{U1lambda1} takes the form
\be
\mu H_\mu -\bar{\mu} {H}_{\bar\mu} + \lambda {H}_\lambda - \bar\lambda {H}_{\bar\lambda} = 0\,.\lb{U1lambda2}
\ee

Using the transformation properties
\be
\delta {\mu} = \alpha -\bar\alpha ({\mu})^2\,, \quad \delta \bar{\mu} = \bar\alpha
- \alpha (\bar{\mu})^2\,,
\ee
it is easy to check that the quantities
\be
\tilde{\mu} = \frac{\mu - \lambda}{1 - \mu\bar\lambda}\,, \qquad \bar{\tilde{\mu}} =
\frac{\bar{\mu} - \bar\lambda}{1 - \bar{\mu}\lambda}\,, \lb{tildemu}
\ee
possess the standard nonlinear realization transformation law
\be
\delta\tilde{\mu} = 2i\tilde{\gamma}\,\tilde{\mu}\,, \quad
\delta\bar{\tilde{\mu}} = -2i\tilde{\gamma}\,\bar{\tilde{\mu}}\,.
\ee
Then we define the Sp(2,$\,\mathbb{R}$) invariant real argument
\be
b := \tilde{\mu}\,\bar{\tilde{\mu}}\,,
\ee
and check that
\be
D_\lambda b  + (\partial_\mu - \bar{\mu}{}^2\partial_{\bar\mu}) b = 0\,, \quad {\rm and \;c.c.}\,.
\ee
This immediately implies that the solution of the constraints \p{alpha2} is
\be
H(\mu, \bar{\mu}, \lambda, \bar\lambda) = I(b)\,.
\ee

Another way to achieve the same result is to make the change of variables $(\mu, \bar{\mu}, \lambda, \bar\lambda)\, \rightarrow \,
(\tilde\mu, \bar{\tilde\mu}, \lambda, \bar\lambda)\,,$ $H(\mu, \bar{\mu}, \lambda, \bar\lambda)
= \tilde{H}(\tilde\mu, \bar{\tilde\mu}, \lambda, \bar\lambda)\,$, in the constraints \p{alpha2} and \p{U1lambda2}. Using the relations
\bea
&& \mu = \frac{\tilde\mu +\lambda}{1 + \tilde\mu\bar\lambda}\,, \qquad 1 + \mu =
\frac{1 + \lambda +\tilde\mu(1 + \bar\lambda)}{1 + \tilde\mu\bar\lambda}\,,\nn
&& H_\mu = \tilde{H}_{\tilde\mu}\, \frac{1 - \lambda\bar\lambda}{(1 - \mu\bar\lambda)^2} =  \tilde{H}_{\tilde\mu}\,
\frac{(1 + \tilde\mu\bar\lambda)^2}{1 - \lambda\bar\lambda}\,,
\eea
one can check that in the new basis the constraints take the form
\bea
&& D_\lambda \tilde{H}  + \bar\lambda (\tilde{\mu} \tilde{H}_{\tilde\mu} -\bar{\tilde\mu} \tilde{H}_{\bar{\tilde\mu}})  = 0\,, \quad
D_{\bar\lambda} \tilde{H}  - \lambda(\tilde{\mu} \tilde{H}_{\tilde\mu} -\bar{\tilde\mu} \tilde{H}_{\bar{\tilde\mu}})= 0\,, \lb{alpha3} \\
&&\tilde{\mu}  \tilde{H}_{\tilde\mu} -\bar{\tilde\mu}  \tilde{H}_{\bar{\tilde\mu}} + \lambda \tilde{H}_\lambda
- \bar\lambda \tilde{H}_{\bar\lambda} = 0\,.\lb{U11lambda2}
\eea
They are equivalent to the set
\be
\tilde{H}_\lambda =  \tilde{H}_{\bar\lambda} =0\,, \quad \tilde{\mu}  \tilde{H}_{\tilde\mu} -\bar{\tilde\mu} \tilde{H}_{\bar{\tilde\mu}} = 0\,,
\ee
whence it follows again that
$$
\tilde{H}(\tilde\mu, \bar{\tilde\mu}, \lambda, \bar\lambda) = I(b)\,.
$$

\subsection{From the $\mu$ representation to the $(F, \bar F)$ Lagrangian}

In the $\mu$ representation, the basic algebraic equation \p{Alggen} implies
\be
\varphi = -H_\mu (1 + \mu)^2\,, \quad \bar\varphi = -H_{\bar\mu} (1 + \bar{\mu})^2\,.\lb{mueq}
\ee
It will be convenient to deal with the variable $\tilde\mu$ defined in \p{tildemu}.
Keeping in mind that for the Sp(2,$\,\mathbb{R}$) invariant case
$$
H_{\tilde\mu} = I'(b) \bar{\tilde\mu}\,, \quad H_{\bar{\tilde\mu}} = I'(b) {\tilde\mu}\,, \quad I'(b) := \frac{d I}{d b }\,,
$$
we can rewrite \p{mueq} as
\be
\varphi = -\frac{1}{1 - \lambda\bar\lambda}\,I'\Big[\bar{\tilde\mu}(1 + \lambda)^2
+ 2 b (1 + \lambda)(1 + \bar\lambda) + \tilde\mu b(1 + \bar\lambda)^2\Big], \quad
\mbox{and c.c.}\,. \lb{mueq1}
\ee
Redefining
\be
\tilde{\mu} = \frac{1 + \lambda}{1 + \bar\lambda}\,\hat\mu\,, \quad \bar{\tilde{\mu}} = \frac{1 + \bar\lambda}{1 + \lambda}\,\bar{\hat\mu}\,, \quad
b = \tilde\mu \bar{\tilde\mu} = \hat{\mu}\bar{\hat\mu} = \hat{b}\,,   \lb{defrho}
\ee
where $\hat\mu$ and $\bar{\hat\mu}$ can be checked to transform just as in \p{hatmutrans}, and introducing
\be
\hat\varphi := \frac{1-\lambda\bar\lambda}{(1 +\lambda)(1 + \bar\lambda)} \,\varphi\,, \quad
\hat{\bar\varphi} = \frac{1-\lambda\bar\lambda}{(1 +\lambda)(1 + \bar\lambda)} \,\bar\varphi\,,\lb{omegadef}
\ee
which, in the $S_1, S_2$ parametrization, coincide with the quantities defined in \p{A8},  we rewrite \p{mueq1} as
\be
\hat\varphi = -(\bar{\hat\mu} + 2 \hat{b} + \hat{b}\hat\mu)\,I'\,, \quad \hat{\bar\varphi} =
-(\hat\mu + 2 \hat{b} + \hat{b}\bar{\hat\mu})\,I'\,,\quad \mbox{and c.c.}\,,\lb{mueq2}
\ee
or
\be
\hat\varphi + \hat{\bar\varphi} + 4I' \hat{b} = -(\hat\mu + \hat{\bar\mu})\,(1 + \hat{b})\,I'\,, \quad \hat\varphi - \hat{\bar\varphi}
= (\hat\mu - \hat{\bar\mu})\,(1 - \hat{b})\,I'\,.\lb{mueq3}
\ee
Eqs. \p{mueq2} are recognized as the basic equation \p{varmu} (and its conjugate) of the $\hat{b}$ representation of the NR formulation presented
in the first part of the paper. So, already at this step we conclude that, on the shell of the auxiliary equation \p{Alggen},  the $b$
representations of both
formulations of the Sp(2,$\,\mathbb{R}$) self-duality are the same, which implies that both these formulations yield the same eventual answer for
the general Sp(2,$\,\mathbb{R}$) self-dual Lagrangian $\tilde{L}^{\rm sd}(\lambda, F)\,$.

It seems instructive to consider here the consequences of eqs. \p{mueq3} in more detail. These equations have the same form
as the equations in the $b$ representation    for the U(1) case without scalars (eqs. (2.37) in \cite{IZ3}),
with the only change $\varphi \rightarrow \hat{\varphi}\,, \bar\varphi \rightarrow \hat{\bar\varphi}\,$ and $\hat\mu, \bar{\hat\mu}$
instead of $\mu, \bar\mu\,$. Hence, as a corollary they imply the same algebraic equation (eq. (2.38) in \cite{IZ3})
\be
(\hat{b} +1)^2 \hat\varphi\hat{\bar\varphi} = \hat{b}\,[\hat\varphi + \hat{\bar\varphi} -(\hat{b}-1)^2I']^2\,, \lb{mueq4}
\ee
which expresses $\hat{b}$ in terms of the variables $\hat\varphi, \hat{\bar\varphi}$ defined in \p{omegadef}. Hence, the solution for $\hat\mu, \bar{\hat\mu}$
is obtained through the substitution
$\varphi\rightarrow \hat\varphi\,, \bar\varphi \rightarrow \hat{\bar\varphi}$ in the solution for $\mu, \bar\mu$ of the case without scalar fields.

It remains to find the general expression for the Lagrangian in the original $(\varphi, \bar\varphi)$ representation. The formulas one starts with
mimic the U(1) case
\be
\tilde{L}^{sd}(\lambda, \varphi, \bar\varphi) = \varphi \tilde{L}^{\rm sd}_\varphi + \bar\varphi \tilde{L}^{\rm sd}_{\bar\varphi}
+ I(\hat{b})\,,\quad \hat{b} =
\hat{\mu}(\lambda, \varphi, \bar\varphi)\bar{\hat\mu}(\lambda, \varphi, \bar\varphi)\,,
\ee
where the $\lambda$ dependence of the r.h.s. is hidden in the $\lambda$ dependence of
$\tilde{L}^{\rm sd}_\varphi, \tilde{L}^{\rm sd}_{\bar\varphi}$ and $\hat{b}$.
The holomorphic derivatives $\tilde{L}^{\rm sd}_\varphi, \tilde{L}^{\rm sd}_{\bar\varphi}$ are related to the variables $\mu, \bar{\mu}$
also by the same relations as in the U(1) case
\be
\tilde{L}^{\rm sd}_\varphi = \frac{\hat{E}_\nu - 1}{2(\hat{E}_\nu + 1)} = \frac{\mu - 1}{2(\mu + 1)}\,, \quad \mbox{and c.c.}\,,
\ee
which, with taking into account eqs. \p{mueq},  yields
\be
\tilde{L}^{\rm sd}= -\sfrac12\,H_\mu\,(\mu^2 - 1)
- \sfrac12\,H_{\bar\mu}\,(\bar{\mu}^2  - 1) + I(b)\,. \lb{lagrhatmu}
\ee
The deviations from the pure U(1) case are revealed, when making use of the basic equations \p{mueq3}, \p{mueq4}
to eliminate $\mu$ and $\bar{\mu}$ in
\p{lagrhatmu}. After passing to the variables $\hat\mu, \bar{\hat\mu}$, we obtain
\be
\tilde{L}^{\rm sd} = \frac12 I' \frac{1 - b}{1 -\lambda\bar\lambda}\Big[ (1 - \lambda\bar\lambda)(\hat\mu + \bar{\hat\mu}) +
(\lambda - \bar\lambda)(\hat\mu - \bar{\hat\mu})\Big] + I(\hat{b})\,.
\ee
Using eqs. \p{mueq3}, we obtain the final expression for the Lagrangian
\bea
&& \tilde{L}^{\rm sd} = \frac12 \frac{\lambda - \bar\lambda}{(1 + \lambda)(1 + \bar\lambda)}(\varphi - \bar\varphi) + \hat{L}(\hat\varphi, \hat{\bar\varphi})\,,
\lb{lagrfin1} \\
&& \hat{L}(\hat\varphi, \hat{\bar\varphi}) = -\sfrac12 (\hat\varphi + \hat{\bar\varphi} + 4\hat{b}I')\,\frac{1 -\hat{b}}{1 + \hat{b}} + I(\hat{b})\,. \lb{tildelagr1}
\eea
The Lagrangian $\hat{L}$ accompanied by the algebraic equation \p{mueq4} precisely yields the standard U(1) self-dual Lagrangian
with the replacements $\varphi \rightarrow \hat{\varphi}, \bar\varphi \rightarrow \hat{\bar\varphi}$, where $\hat\varphi, \hat{\bar\varphi}$
are defined
through $\varphi, \bar\varphi$ by eqs. \p{omegadef} or  \p{A8}. It satisfies the standard GZ self-duality constraint with respect to
$\hat{\varphi}, \hat{\bar\varphi}$:
\be
\hat\varphi - \hat{\bar\varphi} - 4\hat{\varphi}(\hat{L}_{\hat\varphi})^2 + 4 \hat{\bar\varphi} (\hat{L}_{\hat{\bar\varphi}})^2  = 0\,.
\ee
The first term in \p{lagrfin1} is the appropriate modification of the
bilinear part of the action. This final answer for the nonlinear self-dual action is in the precise correspondence with the general action
of Gibbons and Rasheed \cite{GR}. In the $S_1, S_2$ parametrization, using the relations \p{Slambda}, the Lagrangians \p{lagrfin1} and \p{tildelagr1}
can be rewritten in the familiar form as
\bea
&& \tilde{L}^{\rm sd} = -\sfrac{i}2 S_1 (\varphi - \bar\varphi) + \hat{L}(\hat\varphi, \hat{\bar\varphi})\,, \lb{lagrfin} \\
&& \hat{L}(\hat\varphi, \hat{\bar\varphi}) = -\sfrac12 (\hat\varphi + \hat{\bar\varphi} + 4\hat{b}I')\,\frac{1 -\hat{b}}{1 + \hat{b}} + I(\hat{b})\,. \lb{tildelagr}
\eea

Obviously, the Lagrangian \p{lagrfin1} (or \p{lagrfin}) should be accompanied by the coset field Lagrangian  $L'(\lambda) = L(S)$ given in \p{lambdaAct}.
It is straightforward to check that the $\lambda$ equations of motion following from
the total Lagrangian $L^{\rm sd} = L'(\lambda) + \tilde{L}^{\rm sd}$ enjoy  Sp(2,$\,\mathbb{R}$) covariance.

\subsection{Yet another derivation of the Sp(2,$\,\mathbb{R}$) self-dual Lagrangian}

Though the extended Lagrangian \p{LagrVF} contains a constrained auxiliary interaction $E(\nu, \bar\nu, \lambda, \bar\lambda)$
and we cannot immediately solve the relevant constraints \p{U1lambda1} - \p{alpha1}, we know that it yields the correct description of the general
Sp(2,$\,\mathbb{R}$) self-dual systems, as was shown above by passing to its $\mu$ representation. It turns out that the original $E$
formulation is still capable to yield the general Lagrangian $L^{\rm sd}(\lambda, \varphi, \bar\varphi)$ of
the Sp(2,$\,\mathbb{R}$) systems without explicitly solving  \p{U1lambda1} - \p{alpha1}. These constraints amount to
the triplet of alternative  Sp(2,$\,\mathbb{R}$) self-duality constraints on $\tilde{L}^{\rm sd}$ as they given, e.g., in \cite{KT}.

Starting from the Lagrangian \p{LagrVF} and using the auxiliary equation \p{Alggen} together with its corollaries
\be
V\cdot F = \nu (1 + \hat{E}_\nu)\,, \quad \varphi = \nu (1 + \hat{E}_\nu)^2\,,\lb{corol}
\ee
as well as the transformation properties \p{nuTrans1}, \p{tranEnu}, \p{tranE} and
\be
\delta \varphi = 2i\gamma \varphi\, \frac{\hat{E}_\nu -1}{\hat{E}_\nu +1} + 2\alpha \varphi\,\frac{1}{\hat{E}_\nu +1} +
2\bar\alpha \varphi\,\frac{\hat{E}_\nu}{\hat{E}_\nu +1}\,, \lb{deltaphi}
\ee
we find the following simple Sp(2,$\,\mathbb{R}$) transformation of the on-shell (i.e., with the algebraic equation \p{Alggen} taken into account)
Lagrangian:
\be
\delta \tilde{L}^{\rm sd} = i\gamma [\varphi - \bar\varphi - 2(\nu\hat{E}_\nu - \bar\nu\hat{E}_{\bar\nu})] + (\alpha - \bar\alpha)\,(\nu\hat{E}_\nu - \bar\nu\hat{E}_{\bar\nu})\,.
\lb{varL}
\ee

Defining as usual the general dual field strength
\be
P_{\alpha\beta} = i\frac{\partial \tilde{L}^{\rm sd}}{\partial F^{\alpha\beta}} = i(F_{\alpha\beta} - 2V_{\alpha\beta})
\ee
and employing the auxiliary equation \p{Alggen} together with its corollaries \p{corol}, it is easy to show that
\bea
&&F^{\alpha\beta}P_{\alpha\beta} -F^{\dot\alpha\dot\beta}P_{\dot\alpha\dot\beta} = i\nu (\hat{E}_\nu^2 -1) - i\bar\nu (\hat{E}_{\bar\nu}^2 -1)\,, \lb{FP} \\
&& \varphi + P^2 - \bar\varphi - \bar{P}^2 =  4(\nu \hat{E}_\nu - \bar\nu \hat{E}_{\bar \nu})\,. \lb{PP}
\eea
On the other hand, the Lagrangian \p{LagrVF} can be rewritten on the shell of the auxiliary equation as
\be
\tilde{L}^{\rm sd} = \sfrac12\nu(\hat{E}_\nu^2 -1) +  \sfrac12\bar\nu(\hat{E}_{\bar \nu}^2 -1) + H\,,  \quad H =
\hat{E} - \nu\hat{E}_\nu - \bar\nu \hat{E}_{\bar \nu}\,.
\ee
Then, using \p{FP}, we can cast it in the standard form
\be
\tilde{L}^{\rm sd} = \sfrac{i}{2}[(\bar F\cdot \bar P) - (F\cdot  P)] + H\,.  \lb{LagrFP}
\ee
It is straightforward to check that the variation \p{varL} is entirely generated by the first term in \p{LagrFP}, whence it follows that
$\delta H = 0$ in agreement with  \p{defH}.

Substituting the expression \p{PP} into the variation \p{varL}, we can rewrite it in a more standard way in terms
of $\varphi, \bar\varphi, P^2$ and $\bar P^2$. This variation can be also used to find the Sp(2,$\,\mathbb{R}$) analog of the standard
GZ self-duality constraint in the $(F, \bar F)$ representation of the Lagrangian. To this end we can use the relations
of the U(1) self-duality  \cite{IZ2,IZ3}
\be
\nu = \sfrac14\varphi(1 -2 \tilde{L}^{\rm sd}_{\varphi})^2\,, \quad \hat{E}_\nu =
\frac{1 + 2 \tilde{L}^{\rm sd}_{\varphi}}{1 - 2 \tilde{L}^{\rm sd}_{\varphi}}\,,   \lb{nuphi}
\ee
which are valid in the considered case too. We substitute these expressions into the r.h.s. of the variation \p{varL}
and alternatively calculate $\delta \tilde{L}^{\rm sd}$ as
\be
\delta \tilde{L}^{\rm sd} = \delta \lambda \tilde{L}_\lambda^{\rm sd} + \delta\bar\lambda \tilde{L}_{\bar\lambda}^{\rm sd}
+ \delta\varphi \tilde{L}^{\rm sd}_{\varphi}
+ \delta\bar\varphi \tilde{L}^{\rm sd}_{\bar\varphi}\,, \lb{direct}
\ee
where $\delta \varphi$ is defined in \p{deltaphi}. With taking into account \p{nuphi}, the variation \p{deltaphi}  can be rewritten as
\be
\delta \varphi = 4i\gamma\,\tilde{L}_{\varphi}^{\rm sd} + \alpha\,\varphi (1 - 2\tilde{L}_{\varphi}^{\rm sd})
+ \bar\alpha\,\varphi (1 + 2\tilde{L}_{\varphi}^{\rm sd})\,.
\ee
Substituting  the explicit expressions for the variations of $\delta\lambda, \delta\varphi$ and their conjugates
into $\delta \tilde{L}^{\rm sd}$ \p{direct},
and comparing the latter with \p{varL}, we obtain three conditions on the Lagrangian $\tilde{L}^{\rm sd}$ which are just the Sp(2,$\,\mathbb{R}$)
extension of the standard GZ constraint:
\bea
&&4(\lambda \tilde{L}_\lambda^{\rm sd} - \bar\lambda \tilde{L}_{\bar\lambda}^{\rm sd}) = \varphi[1 - 4(\tilde{L}_{\varphi}^{\rm sd})^2] -
\bar\varphi[1 - 4(\tilde{L}_{\bar\varphi}^{\rm sd})^2]\,, \lb{one} \\
&&4D_\lambda \tilde{L}^{\rm sd} =  \varphi(1 - 2\tilde{L}_{\varphi}^{\rm sd})^2 - \bar\varphi(1 + 2\tilde{L}_{\bar\varphi}^{\rm sd})^2\,, \lb{two} \\
&& 4D_{\bar\lambda} \tilde{L}^{\rm sd} = \bar\varphi(1 - 2\tilde{L}_{\bar\varphi}^{\rm sd})^2 - \varphi(1 + 2\tilde{L}_{\varphi}^{\rm sd})^2\,, \lb{three}
\eea
where $D_{\lambda, \bar\lambda}$ were defined in \p{defDlambda}. These constraints are equivalent to the sets \p{U1lambda1} - \p{alpha1}
or \p{alpha2}, \p{U1lambda2}. One can explicitly check that the Lagrangian in the form \p{lagrfin}, \p{tildelagr} solves eqs. \p{one} - \p{three}.
They can be cast in a more familiar form after passing to the coset representatives $S_1,S_2$ by the formulas \p{Slambda}.
In the new parametrization, the $\lambda$ derivatives are expressed as
\bea
&& \lambda\partial_\lambda - \bar\lambda\partial_{\bar\lambda} = \sfrac{i}{2}(\partial_{S} +  \partial_{\bar{S}} + {S}^2\partial_{S} +
\bar{S}^2\partial_{\bar{S}})\,, \nn
&& D_\lambda +  D_{\bar\lambda} = -2( {S}\partial_{S} + \bar{S}\partial_{\bar{S}})\,, \nn
&& D_\lambda -  D_{\bar\lambda} = i(\partial_{S}+  \partial_{\bar{S}} - {S}^2\partial_{S} - \bar{S}^2\partial_{\bar{S}})\,.
\eea
Using these relations and going over to the tensorial notation:
\bea
&& \varphi \tilde{L}_{\varphi}^{\rm sd} + \bar\varphi \tilde{L}_{\bar\varphi}^{\rm sd} = \sfrac12 {F}^{mn}\frac{\partial \tilde{L}^{\rm sd}}{\partial {F}^{mn}}\,, \quad
i(\varphi - \bar\varphi) = -\sfrac12 {F}^{mn}\tilde{{F}}_{mn}\nn
&& i[\varphi (\tilde{L}^{\rm sd}_{\varphi})^2 - \bar\varphi (\tilde{L}^{\rm sd}_{\bar\varphi})^2] = -\sfrac{i}{4}(P^2 - \bar{P}^2) =
\sfrac18 {G}^{mn}\tilde{{G}}_{mn}\,,
\eea
we can bring  \p{one} - \p{three} to the simple equivalent form given in \cite{KT}
\bea
&& 2({S}\tilde{L}^{\rm sd}_{S} + \bar{S}\tilde{L}^{\rm sd}_{\bar{S}}) = {F}^{mn}\frac{\partial \tilde{L}^{\rm sd}}{\partial {F}^{mn}}\,, \nn
&& \tilde{L}^{\rm sd}_{S} + \tilde{L}^{\rm sd}_{\bar{S}}  = \sfrac14 {F}^{mn}\tilde{{F}}_{mn}\,, \nn
&& {S}^2\tilde{L}^{\rm sd}_{S} +  \bar{S}^2 \tilde{L}^{\rm sd}_{\bar{S}} =  \sfrac14 {G}^{mn}\tilde{{G}}_{mn}\,. \lb{kt}
\eea
The unique solution of these constraints is the general GR Lagrangian \p{Sp2Rlagr}, \p{A7} (with $L(S)$ subtracted)\footnote{Note
that the constraints \p{one} - \p{three} or \p{kt} are formulated for the Lagrangian $\tilde{L}^{\rm sd} =L^{\rm sd} - L(S)\,$, while the earlier employed the
GZ-type  constraints \p{A9} or \p{A17} only for its part $\hat{L}(\hat{\varphi}, \hat{\bar\varphi})$. These two sets of constraints
are in fact equivalent to each other.}. So  the linear realization version of the  bispinor auxiliary field
formulation of the Sp(2,$\,\mathbb{R}$) self-dual systems  yields as the output the same GR Lagrangian as the formulation based
on the nonlinearly transforming auxiliary fields, even without passing to the $\mu$ representation.

\subsection{More on the interplay between the linear and nonlinear formulations}

Here we give more details on the relationship between the auxiliary fields in the LR and NR formulations.

In Subsections 4.2 and 4.3 we observed that the Legendre transformation from the variables $\nu, \bar\nu$ to $\mu, \bar\mu$ performed in \p{LagrVF}
 yields in fact the same $\mu$ representation  as the Legendre  transformation  from the variables $\hat\nu, \hat{\bar\nu}$ to $\hat\mu, \hat{\bar\mu}$
in the Lagrangian \p{calLtot1}. It is interesting to reproduce the NR formulation, starting  from the $\mu$ representation obtained
in the framework of the LR formulation and applying just another type of the inverse Legendre transformation
to this $\mu$ representation.

Namely, we start from the Sp(2,$\,\mathbb{R}$) invariant function $H(\mu, \bar\mu, \lambda, \bar\lambda)$ defined in \p{defHmu},
make the equivalency transformation from the variables $\mu, \bar\mu$ to $\hat\mu, \hat{\bar\mu}$ according to
$$
\mu = \frac{(1 + \lambda)\hat\mu + \lambda(1+\bar\lambda)}{1 + \bar\lambda + (1 + \lambda)\bar\lambda\hat\mu}\,,
$$
where the relations \p{tildemu} and \p{defrho} were used, define
\bea
H(\mu, \bar\mu, \lambda, \bar\lambda) = \hat{H}(\hat\mu, \hat{\bar\mu}, \lambda, \bar\lambda) \lb{hatH}
\eea
and perform the Legendre transformation with respect to the variables $\hat\mu, \hat{\bar\mu}$:
\bea
\hat{H}_{\hat{\mu}}  := - \hat{\nu}\,, \;\; \hat{H}_{\hat{\bar\mu}} = -\bar{\hat\nu}\,, \quad
\hat{E} = \hat{H} - \hat\mu\hat{H}_{\hat{\mu}} -  \hat{\bar\mu} \hat{H}_{\hat{\bar\mu}}\,.
\eea
Because $\delta \hat{H} = 0$, we immediately find
\bea
\delta \hat{\nu} = -(2i\gamma - \alpha + \bar\alpha)\,{S}_2\, \hat{\nu} = -2i \rho \hat{\nu}\,, \quad  \mbox{and c.c.}\,,  \lb{deltanutil}
\eea
and
\bea
\delta \hat{E} = 0\,,
\eea
in the full agreement with the basic formulas of the NR representation. We also obtain
\be
\hat\mu =  \hat{E}_{\hat\nu}\,, \quad \hat{\bar\mu} = \hat{E}_{\bar{\hat\nu}}\,, \quad  \hat{H} = \hat{E}
- \hat{\nu}\hat{E}_{\hat{\nu}} -  \hat{\bar\nu} \hat{E}_{\hat{\bar{\nu}}}\,.
\ee
The equation
\be
\hat\mu \hat{H}_{\hat\mu} - \hat{\bar\mu}\hat{H}_{\hat{\bar\mu}} = 0
\ee
implies
\be
\hat{\nu} \hat{E}_{\hat{\nu}}  - \hat{\bar{\nu}}\hat{H}_{\hat{\bar{\nu}}} = 0 \quad \Rightarrow \quad
\hat{E} = {\cal E}(\hat{a})\,, \; \hat{a} = \hat{\nu}\bar{\hat{\nu}}\,.
\ee
Eq. \p{mueq2} acquires the following form in terms of the newly introduced variables
\be
\hat\phi = \hat{\nu}( 1 + \hat{E}_{\hat\nu})^2\,.  \lb{omega2}
\ee
Using the transformation laws of the variables $\hat\mu$, $\hat\nu$ and $\hat\phi$ it is straightforward
to check the Sp(2,$\,\mathbb{R}$) covariance of \p{omega2}.

In this way the basic formulas of the NR formulation are recovered.

As for the  Lagrangian \p{calLtot1}, it can be restored from the requirement that it reproduces the basic relations of the NR formulation,
including eq. \p{omega2}. We introduce the bispinor  field $\hat{V}_{\alpha\beta}, \bar{\hat{V}}_{\dot\alpha\dot\beta}$, such that
\be
\hat\nu = \hat{V}^{\alpha\beta}\hat{V}_{\alpha\beta}\,, \quad  \hat{\bar\nu} = \hat{\bar{V}}^{\dot\alpha\dot\beta}\hat{\bar{V}}_{\dot\alpha\dot\beta}\,,
\ee
and represent the sought Lagrangian as
\be
{\cal L}(\lambda, F, \hat{V}) = {\cal L}_1(\lambda, \varphi) - 2 \sqrt{S_2}[(\hat{V}\cdot  F) +
(\hat{\bar{V}}\cdot\bar F)] + \hat{\nu} + \hat{\bar{\nu}} + {\cal E}(\hat\nu\hat{\bar{\nu}})\,. \lb{Ltilde}
\ee
Varying it with respect to $\hat{V}_{\alpha\beta}$ gives
\be
\sqrt{S_2}\,F_{\alpha\beta} = (1 + \hat{E}_{\hat{\nu}})\,\hat{V}_{\alpha\beta}\,, \lb{tildeVF}
\ee
which implies just \p{omega2}. We also need the correct dynamical equations for $F_{\alpha\beta}$ on the shell of the algebraic constraint.
In other words, we require
\be
F_{\alpha\beta} - 2V_{\alpha\beta} = \frac{\partial {\cal L}}{\partial F^{\alpha\beta}} =
\left( \frac{\partial {\cal L}_1}{\partial F^{\alpha\beta}} -
2 \sqrt{S_2}\,\hat{V}_{\alpha\beta}\right),  \lb{FVcond}
\ee
with $\hat{V}_{\alpha\beta}$ being subjected to \p{tildeVF}.

As the first step, we rewrite $F_{\alpha\beta} - 2V_{\alpha\beta}$ in terms of the variables $\hat\nu$
\be
F_{\alpha\beta} - 2V_{\alpha\beta} = \Big(-i S_1 + S_2\frac{\hat{E}_{\hat{\nu}} - 1}{\hat{E}_{\hat{\nu}} + 1} \Big)F_{\alpha\beta}. \lb{auxFin1}
\ee
Next we express $\hat{V}_{\alpha\beta}$ in \p{FVcond} through $F_{\alpha\beta}$ by  \p{tildeVF} and compare two expressions
for  $F_{\alpha\beta} - 2V_{\alpha\beta}\,$, which yields
\be
{\cal L}_1 = \sfrac12 {S}_2\,(\varphi + \bar\varphi)- \sfrac{i}{2} {S}_1\,(\varphi -\bar\varphi)\,.
\ee

Finally, the Lagrangian \p{Ltilde} takes the form of \p{calLtot1} with the subtracted $L(S)$. Note that the auxiliary equation \p{tildeVF}
together with \p{auxFin1} yield the same on-shell relations \p{VhatV} ad \p{nuhatnu} between the LR and NR tensorial fields.

The off-shell Lagrangians \p{LagrVF} and \p{calLtot1} are essentially different and seem not to be related
to each other by any obvious field redefinition, though they yield the same system on the shell of the auxiliary equation.

\section{Conclusions}

We investigated Sp(2,$\,\mathbb{R}$) duality-invariant interactions
of scalar and electromagnetic fields, employing two different
formulations involving auxiliary bispinors and/or auxiliary scalars (``$\mu$ representation'').
The main emphasis was on the transformation properties of the relevant Lagrangians
and their equations of motion. The formalism of Section 3 started from the nonlinear
realization of Sp(2,$\,\mathbb{R}$) on the basic auxiliary bispinor fields, while
in Section 4 those auxiliary fields were taken to transform linearly.
Both formalisms admit a Legendre-type transformation to
Lagrangians with auxiliary scalar fields. This allowed us to prove that both auxiliary-field
formulations yield equivalent self-dual Lagrangians in the standard $(S,F)$ representation.
Like in the U(1) duality case, any choice of a Lagrangian
exhibiting Sp(2,$\,\mathbb{R}$) duality amounts to a particular choice of an
Sp(2,$\,\mathbb{R}$) invariant unconstrained interaction of the auxiliary bispinor fields.
The $(S,F)$ Lagrangian emerges from the extended Lagrangian upon elimination of the auxiliary fields
through their equations of motion in terms of  the coset scalars $S$ and the electromagnetic field
strengths $F$.

It is rather straightforward to generalize our auxiliary-field formulations to the case of
Sp($2N,\,\mathbb{R}$) duality as proper extensions of the analogous formulations
for U($N$) duality \cite{IZ4}. We hope to address this task elsewhere.
The formalism of auxiliary superfields in ${\cal N}=1$ supersymmetric self-dual theories \cite{Ku, ILZ}
can also be generalized, with additional chiral coset multiplets, to  the case of noncompact dualities.
We are curious to learn which of the two (if not both) approaches presented here
admit an unambiguous extension to supersymmetric theories.
One might also try to construct an Sp(2,$\,\mathbb{R}$) (and Sp($2N,\,\mathbb{R}$)) version
of the ``hybrid'' formulation of U(1) duality \cite{hybrid}, which joins the auxiliary-tensor approach
with the manifestly Lorentz- and duality-invariant PST formalism
(see \cite{PST} and references therein).

\section*{Acknowledgements}
We acknowledge partial support from the RFBR grants No 12-02-00517,
No 13-02-90430, No 13-02-91330, the grant DFG LE 838/12-1 and a
grant of the Heisenberg-Landau program. E.I. and B.Z. thank the
Institute of Theoretical Physics at Leibniz University Hannover for
kind hospitality during several visits in the course of this work.


\begin{thebibliography}{99}
\addtolength{\itemsep}{-3pt}

\bibitem{GZ}M.K. Gaillard and B. Zumino, {\it Duality rotations for interacting
fields}, \\ Nucl. Phys. B {\bf 193} (1981) 221;\\
M.K. Gaillard and B. Zumino, {\it Self-duality in nonlinear
electromagnetism}, \\ in: Supersymmetry and quantum field theory, eds.
J. Wess and V.P. Akulov, p. 121-129,\\
Springer-Verlag, 1998; {\tt arXiv:hep-th/9705226};\\
M.K. Gaillard and B. Zumino, {\it Nonlinear electromagnetic self-duality
and Legendre transform},\\ in: Duality and supersymmetric theories,
eds. D.I. Olive and P.C. West, p. 33, \\ Cambridge University Press,
1999; {\tt arXiv:hep-th/9712103}.
\bibitem{GR} G.W. Gibbons and D.A. Rasheed, \\ {\it Electric-magnetic duality
rotations in nonlinear electrodynamics}, \\ Nucl. Phys. B {\bf 454} (1995)
185; {\tt arXiv:hep-th/9506035}.
\bibitem{GR2} G.W. Gibbons and D.A. Rasheed, \\ {\it $SL(2,R)$ invariance of nonlinear
electrodynamics coupled to axion and dilaton}, \\ Phys. Lett. B {\bf 365}
(1996) 46; {\tt arXiv:hep-th/9509141}.
\bibitem{KT} S.M. Kuzenko and S. Theisen, {\it Nonlinear self-duality and
supersymmetry}, \\ Fortsch. Phys. {\bf 49} (2001) 273; {\tt arXiv:hep-th/0007231}.
\bibitem{AFZ} P. Aschieri, S. Ferrara and B. Zumino,\\ {\it Duality
rotations in nonlinear electrodynamics and extended supergravity},\\
Riv. Nuovo Cim. {\bf 31} (2008) 625; {\tt arXiv:0807.4039 [hep-th]}.
\bibitem{Sgrav} D.Z. Freedman and A. Van Proeyen, {\it Supergravity},
Cambridge University Press, 2012, 607p.
\bibitem{IZ} E.A. Ivanov, B.M. Zupnik, {\it N=3 supersymmetric Born-Infeld
theory}, \\ Nucl. Phys. B {\bf 618} (2001) 3; {\tt arXiv:hep-th/0110074}.
\bibitem{IZ1} E.A. Ivanov, B.M. Zupnik, \\ {\it New representation for Lagrangians
 of self-dual nonlinear electrodynamics}, \\ in: Supersymmetries and
quantum symmetries, eds. E. Ivanov et al, p. 235, Dubna, 2002; \\
{\tt arXiv:hep-th/0202203}.
\bibitem{IZ2}
 E.A. Ivanov, B.M. Zupnik, \\ {\it New approach to nonlinear
electrodynamics: dualities as symmetries of interaction}, \\ Yadern.
Fiz.  {\bf 67} (2004)  2212 [Phys.  Atom.
Nucl.  {\bf 67} (2004) 2188]; {\tt arXiv:hep-th/0303192}.
 \bibitem{IZ3}
 E.A. Ivanov, B.M. Zupnik,\\ {\it Bispinor auxiliary fields in duality-invariant electrodynamics
 revisited}, \\ Phys. Rev. D {\bf 87} (2013) 065023; {\tt arXiv:1212.6637 [hep-th]}.
\bibitem{IZ4}
 E.A. Ivanov, B.M. Zupnik, \\ {\it Bispinor auxiliary fields in duality-invariant electrodynamics
 revisited, the $U(N)$ case}, \\ Phys. Rev. D {\bf 88} (2013)  045002; {\tt arXiv:1304.1366 [hep-th]}.
\bibitem{Ku}S.M. Kuzenko, {\it Duality rotations in supersymmetric nonlinear
 electrodynamics revisited},  \\ JHEP {\bf 1303} (2013) 153; {\tt arXiv:1301.5194 [hep-th]}.
\bibitem{ILZ} E. Ivanov, O. Lechtenfeld,  B. Zupnik, \\ {\it Auxiliary superfields
 in  ${\cal N}=1$  supersymmetric self-dual electrodynamics}, \\ JHEP {\bf 1305}
 (2013) 133; {\tt arXiv:1303.5962 [hep-th]};\\
 E. Ivanov, O. Lechtenfeld,  B. Zupnik,\\ {\it New approach to duality-invariant
nonlinear electrodynamics}, \\ Jour.  Phys. Conf. Ser. {\bf 474} (2013)
012023; {\tt arXiv:1310.5362 [hep-th]}.
 \bibitem{IZ5} E.A. Ivanov, B.M. Zupnik,
 {\it Self-dual ${\cal N}=2$ Born-Infeld theory through auxiliary superfields},\\
 JHEP {\bf 1405} (2014) 061; {\tt arXiv:1312.5687 [hep-th]}.
\bibitem{BN} G. Bossard, H. Nicolai, {\it Counterterms vs. dualities},\\
JHEP {\bf 1108} (2011) 074; {\tt arXiv:1105.1273 [hep-th]}.
\bibitem{CKR} J.J.M. Carrasco, R. Kallosh, R. Roiban, \\ {\it Covariant procedure for
perturbative nonlinear deformation of duality-invariant theories},\\
Phys. Rev. D {\bf 85} (2012) 025007; {\tt arXiv:1108.4390 [hep-th]}.
\bibitem{CKO}W. Chemissany, R. Kallosh, T. Ortin, {\it Born-Infeld with higher derivatives},\\
Phys. Rev. D  {\bf 85} (2012) 046002; {\tt arXiv:1112.0332 [hep-th]}.
\bibitem{Nonl} S.R. Coleman, J. Wess, B. Zumino, \\
{\it Structure of phenomenological Lagrangians. 1.},
Phys. Rev. {\bf 177} (1969) 2239; \\
C.G. Callan, Jr., S.R. Coleman, J. Wess, B. Zumino,\\
{\it Structure of phenomenological Lagrangians. 2.},
Phys. Rev. {\bf 177} (1969) 2247.
\bibitem{hybrid} E.A. Ivanov, A.J. Nurmagambetov, B.M. Zupnik, \\
{\it Unifying the PST and the auxiliary tensor field formulations of 4D self-duality},\\
Phys. Lett. B {\bf 731} (2014) 298; {\tt arXiv:1401.7834 [hep-th]}.
\bibitem{PST} P. Pasti, D. Sorokin and M. Tonin, \\ {\it Covariant actions for models
with non-linear twisted self-duality},  \\ Phys. Rev. D {\bf 86} (2012)
045013; {\tt arXiv:1205.4243 [hep-th]}.

\end{thebibliography}
\end{document}